\documentclass{article}

\usepackage[utf8]{inputenc}
\usepackage[margin=0.75in]{geometry}
\usepackage{amsmath,amssymb,amsfonts,amsthm}
\usepackage{bm}
\usepackage{graphicx}
\usepackage{booktabs}
\usepackage{multirow}
\usepackage{longtable}
\usepackage{subcaption}
\usepackage{pdflscape} 
\usepackage{titling}   
\usepackage{hyperref}
\usepackage{xcolor}
\usepackage{cite}
\usepackage{cleveref}

\hypersetup{
    colorlinks=true,
    linkcolor=blue,
    citecolor=blue,
    urlcolor=blue
}

\newcommand{\real}{\mathbb{R}}

\newcommand{\mat}[1]{\mathbf{#1}}

\title{\textbf{A Multi-Model Non-Intrusive Reduced-Order Framework for Parametric Erosion Prediction via Kinematic Cross-Moment Compression}}

\author{\textbf{Animesh Yadav$^{1}$, Rajesh Kumar Shukla$^{1, *}$, Ravinder Kumar Duvedi$^{1}$}\\[0.25em]
\small{$^{*}$Corresponding author}\\[0.5em]
\small\textit{$^{1}$Department of Mechanical Engineering, Thapar Institute of Engineering and Technology, Patiala, Punjab 147004, India}}
\date{}

\begin{document}

\maketitle

\begin{abstract}
High-fidelity Eulerian--Lagrangian simulations of solid particle erosion in curved pipes require hours of compute per operating point, preventing rapid parameter sweeps and real-time wear assessment. Existing reduced-order models (ROMs) speed up these evaluations, yet they are typically trained on a single, fixed empirical erosion formula (e.g., Oka or Finnie). Changing the material law or target hardness then requires a complete retrain of the surrogate. Here, we present a non-intrusive reduced-order framework that avoids this model-locking by approximating the underlying particle collision kinematics instead of scalar wear rates. Specifically, we project and compress 23 Eulerian boundary cross-moments ($\mathbb{E}[V_p^u \sin^v\alpha_p \cos^w\alpha_p]$) across the pipe surface.

Using 372 high-fidelity CFD-DPM cases of $90^\circ$ elbows over three bend ratios ($R/D \in \{1.5, 2.0, 5.0\}$), five Reynolds numbers, five density ratios, and six particle diameters in the inertial regime ($St > 1$), we evaluate a hybrid compression scheme. Linear Proper Orthogonal Decomposition (POD) and Mode-1 tensor unfolding SVD are combined with block-wise Convolutional Autoencoders (CNN-AE) to handle both broad convective transport and localized impact craters. An anisotropic Gaussian Process Regression (GPR) surrogate maps four dimensionless $\Pi$-groups to the compressed latent space, evaluating full 2D wear topographies in roughly $2\,\mathrm{ms}$ ($R^2 > 0.99$ on primary kinematic fields). Because kinematics are decoupled from material damage laws, the resulting surrogate evaluates multiple empirical models post-hoc exactly matching Finnie and closely approximating Oka, McLaury, and Arabnejad without retraining.
\end{abstract}


\section{Introduction}
\label{sec:introduction}

Solid particle erosion in slurry and pneumatic transport systems represents a critical point of failure across energy, resource, process and manufacturing industries \cite{parsi2014review,kang2024erosive}. In oil and gas extraction and sub-sea transport pipelines, entrained sand particles cause wall thinning in the pipeline which leads catastrophic blowouts \cite{parsi2014review,zahedi2018,othman2024integration,chen2024novel} causing economic losses and environmental degradation. In mineral processing, dredging conduits, and hydraulic transport networks, abrasive slurries erode the pipeline infrastructure \cite{sinha2017review,espinozajara2025,clark1992,desale2009}. In thermal power plants, fly-ash conveying lines experience structural degradation \cite{kang2024erosive,wang2022numerical}. In advanced manufacturing, abrasive particulate flows are actively harnessed in Abrasive Jet Machining (AJM) and Abrasive Water/Slurry Jet Polishing (AJP/AWJP) for high-precision optical and medical component surface finishing \cite{chen2017ajp}. In all these piping circuits, bent pipe sections (elbows) are geometrically indispensable to achieve compact equipment footprints; however, they consistently experience the most aggressive, localized wall thinning and premature catastrophic failure across the entire conveying network \cite{parsi2014review,sinha2017review,mahmoudi2025erosion}.


Erosive wear in pipe bends is driven by the interaction of carrier fluid dynamics, secondary Dean vortex circulations, and dispersed particulate transport \cite{clark1992,desale2009,kang2024erosive,espinozajara2025}. As the carrier flow traverses the bend, the radial pressure gradient drives secondary cross-stream circulation that produces a pair of counter-rotating Dean vortices onto the primary velocity field. At the microscopic boundary interface, material degradation occurs due to competing micromechanical mechanisms: acute grazing impactions ($\alpha < 30^\circ$) induce micro-cutting and directional ploughing, whereas steep normal collisions ($\alpha \approx 90^\circ$) cause repeated plastic deformation, subsurface work hardening, and surface fatigue \cite{finnie1960erosion,bitter1963,bellman1981erosion,albukhaiti2007}. Particle transport regimes are largely determined by the particle Stokes number $St = \tau_p / \tau_f = (\rho_p d_p^2 U_{\mathrm{in}})/(18 \mu_f D)$ \cite{hajisharifi2023non,mahmoudi2025erosion}:
\begin{itemize}
    \item \textbf{Low-Inertia Regime ($St < 0.1$):} Fine particles follow the carrier fluid streamlines and secondary Dean vortices, causing diffuse, sweep-induced wear along the lateral sidewall crowns ($\theta \approx \pm \pi/2$) \cite{espinozajara2025}. We found that the magnitude of erosion in this regime was almost negligible (refer Section \ref{subsec:stokes_filtering}) as compared to the High-Inertia regime.
    \item \textbf{High-Inertia Regime ($St \gg 1$):} In this regime, particle inertia far exceeds drag, so the heavier particles cut across carrier streamlines and strike the outer extrados wall ($\theta = 0$) as well as the downstream reattachment zone, producing deep, localized wear craters \cite{clark1992,chen2004cfd,mahmoudi2025erosion}.
\end{itemize}

The study of solid particle erosion has its roots in semi-empirical wear equations. Finnie \cite{finnie1960erosion} developed one of the earliest ductile micro-machining models, and Bitter \cite{bitter1963} later incorporated both cutting and plastic deformation mechanisms. These foundational formulations generally treated erosion as a power-law in impact velocity ($E \propto V^n$, $n \approx 1.7$--$2.6$) and particle diameter ($d_p^m$) \cite{clark1992,elkholy1983prediction,desale2009}. Subsequent work expanded the scope to include target hardness, particle shape, and the non-monotonic dependence of wear on impact angle for various alloys, eventually producing the widely adopted engineering models of McLaury and Shirazi \cite{mclaury1996,mclaury1999,chen2004cfd}, Ahlert \cite{ahlert1994}, Oka et al.~\cite{oka2005erosion1,oka2005erosion2}, and Arabnejad et al.~\cite{arabnejad2015}

With the advent of mature CFD solvers, the Eulerian--Lagrangian (CFD--DPM) framework became the workhorse for resolving 3D erosion patterns in piping systems \cite{chen2004cfd,peng2016numerical,parsi2017cfd,singh2019modeling,wee2019cfd,wang2021experimental,mahmoudi2025erosion}. The fluid phase is solved on an Eulerian grid---usually with Reynolds-Averaged Navier--Stokes (RANS) or Large Eddy Simulation (LES) \cite{launder1972lectures} while individual particle paths are integrated in Lagrangian coordinates, often requiring hundreds of thousands of trajectories for statistical convergence. Turbulent dispersion is typically handled through Discrete Random Walk (DRW) models, paired with empirical wall restitution coefficients such as those of Grant and Tabakoff \cite{grant1975} and Haider et al. \cite{Haider2017Mechanistic}. Benchmark comparisons by Solnordal et al.~\cite{solnordal2015} and Parsi et al.~\cite{parsi2014review} confirm that CFD--DPM reliably captures both the position and depth of the primary wear scar in $90^\circ$ elbows. The practical bottleneck is compute time: a single 3D simulation routinely requires $6$--$8$ CPU hours, which rules out direct CFD for real-time structural monitoring, remaining-life assessment, or broad parametric design sweeps \cite{wang2022numerical,mondal2022burst}.

This computational expense has motivated a growing body of work on machine learning (ML) surrogates and reduced-order models (ROMs) for erosion (see Jindal et al.~\cite{jindal2026artificial} {for a recent review}). Initial efforts targeted scalar wear-rate prediction, employing Random Forest regression \cite{zahedi2018,yang2021experimental}, swarm-optimized algorithms, gradient-boosted trees \cite{wang2022solid,zhang2021solid}, and Extreme Learning Machines \cite{wang2022numerical,liu2021exploration}. Attention has since shifted toward predicting the full spatial wear distribution. Pandya et al.~\cite{pandya2020cfd} and Othman et al.~\cite{othman2024integration} trained Artificial Neural Networks (ANN) on CFD data, whereas Ma et al.~\cite{ma2024mechanism}, Chen et al.~\cite{chen2024novel}, and Shojaie et al.~\cite{shojaie2023method,shojaie2024comprehensive} built physics-guided Gaussian Process Regression (GPR) models that additionally furnish uncertainty estimates. On the deep-learning side, Li et al.~\cite{li2026cfd} recently trained a CNN--LSTM network, optimized with a Coati Optimization Algorithm (COA), to predict transient wear in slurry elbows.

At the same time, spatial projection techniques like Proper Orthogonal Decomposition (POD) \cite{sirovich1987turbulence,rozza2008reduced,quarteroni2015reduced}, tensor decompositions (HOSVD/Tucker) \cite{tucker1966some,lathauwer2000multilinear,barragan2025motif,alsayyari2019}, and convolutional autoencoders (CNN-AE) \cite{fresca2021pod,xiao2019,murata2020nonlinear,hajisharifi2023non,jiang2023data,garcia2016data} have become very popular for speeding up single-phase CFD and other PDE systems \cite{brunton2019data}. But despite all this progress in CFD and ML surrogates, there are still a few major problems with how the community currently models erosion:

\begin{enumerate}
    \item \textbf{Zero-Dimensional (0D) Approach:} Most existing ML approaches \cite{zahedi2018,othman2024integration,wang2022numerical,wang2022solid,zhang2021solid} compress the complex erosion field into a single scalar (e.g. global maximum penetration rate, total mass lost, etc.). This method disregards the entire spatial wear topography, circumferential crater width, and downstream secondary patterns which are necessary for structural integrity assessment.

    \item \textbf{"Model-Locking":} Existing spatial erosion surrogates in literature are "model-locked" \cite{ma2024mechanism,chen2024novel,li2026cfd,pandya2020cfd}, i.e. the ROMs are trained using a specific empirical erosion equation (e.g., Oka or DNV) \cite{finnie1960erosion,mclaury1996,oka2005erosion1,oka2005erosion2,arabnejad2015}. Since the model constants for empirical wear relations are dependent on metallurgy, surface hardness, particle shape, and material interaction \cite{parsi2014review,kang2024erosive,jindal2026artificial}, changing even one of these parameters requires re-running the entire CFD-DPM simulation and re-training the ROMs.

    \item \textbf{Spatial Sparsity and High-Inertia Craters:}  Wear in elbows is heavily localized: over $70\%$ of the pipe surface (the intrados) receives practically zero particle collisions, while damage concentrates in a small patch on the extrados ($\sim 30\%$ area). Standard unweighted POD and monolithic autoencoders minimize global $L_2$ error, causing them to fit the broad zero-wear regions while underpredicting sharp peak depths in heavy ballistic flows ($St \gg 1$).

\end{enumerate}

To address these fundamental barrier, this paper presents a NI-ROM framework for parametric erosion prediction in pipe bends. By abstracting particulate impact mechanics into fundamental kinematic cross-moments ($\mathbb{E}[V^u f_v(\alpha)]$), our framework completely decouples the spatial reduced-order representation from specific empirical erosion formulas. This enables instantaneous, post-hoc evaluation of arbitrary linear and non-linear wear models (Oka, Finnie, McLaury, Arabnejad, DNV) across diverse pipeline substrates without requiring any surrogate retraining.

\textbf{Representable Function Class and Approximation Error}

\begin{table}[htpb]
\centering
\caption{Classification of erosion formulations and their representability within the selected 23-moment kinematic basis.}
\label{tab:representability}
\small
\resizebox{\linewidth}{!}{%
\begin{tabular}{lllll}
\toprule
\textbf{Erosion Model} & \textbf{Velocity Dependence} & \textbf{Angular Dependence} & \textbf{Exactly Representable?} & \textbf{Approximation Required?} \\
\midrule
Finnie & $V^2$ & $\sin(2\alpha), \cos^2\alpha$ & Yes & No \\
Oka & $V^{2.338}$ & $f(\alpha)$ non-polynomial & No (Approximated via basis) & Yes (Velocity \& Angle) \\
McLaury & $V^{1.73}$ (mapped to $V^2, V^1$) & $f(\alpha)$ piecewise poly & No (Approximated via basis) & Yes (Velocity \& Angle) \\
Arabnejad & $V^{2.41}, V^2$ & $f(\alpha)$ piecewise cutting/def. & No (Approximated via basis) & Yes (Velocity \& Angle) \\
Withheld Model & $V^{2.338}$ & $\sin^{1.5}\alpha (1-\cos\alpha)^{0.5}$ & No & Yes \\
\bottomrule
\end{tabular}%
}
\end{table}

The 23-moment basis can exactly reproduce any erosion formula whose functional form lies within the span of these retained polynomials. For models that fall outside this span, we project them onto the basis, which inevitably introduces a representation error:
\begin{equation}
    \mathbb{E}[g(V_p, \alpha_p)] = \sum_{k=1}^{23} c_k M_k, \quad \text{where } M_k = \mathbb{E}[V_p^{u_k} \sin^{v_k}\alpha_p \cos^{w_k}\alpha_p]
    \label{eq:exact_representation}
\end{equation}
This exact representation holds for polynomial-separable models $g(V,\alpha) = \sum_{u,v,w} A_{u,v,w} V^u \sin^v\alpha \cos^w\alpha$ with velocity powers $u \in \{1, 2, 2.5, 3\}$ and trigonometric powers $v,w \in \{0,1,2,3\}$, which includes Finnie. Formulations with non-polynomial angular functions or fractional exponents (e.g., Oka, McLaury, and Arabnejad) are projected onto the polynomial basis prior to taking the expectation over the particle ensemble:
\begin{equation}
    \mathbb{E}[g(V_p, \alpha_p)] \approx \sum_{k=1}^{23} c_k M_k + \epsilon_g
    \label{eq:approx_representation}
\end{equation}
Here $\epsilon_g$ is the truncation error from terms not spanned by the basis. In practice, the 23 moments were chosen by analyzing the velocity and angular dependencies in the Oka, Finnie, McLaury, and Arabnejad equations, augmented with higher powers ($V^3$, $\sin^3\alpha$) to ensure broad coverage.

The primary contributions of this work are:
\begin{enumerate}
    \item[(i)] \textbf{Kinematic Cross-Moment Abstraction:} We introduce a functional-basis formulation that represents particle wall impacts via 23 Eulerian cross-moments spanning velocity powers ($V^u$, $u \in \{1,2,2.5,3\}$), impact angle terms ($\sin^v\alpha$, $\cos^w\alpha$), and collision frequencies. This permits rapid post-hoc evaluation of diverse wear laws (exact for Finnie, approximate for Oka, McLaury, Arabnejad) without surrogate retraining.
    
    \item[(ii)] \textbf{Non-Intrusive Spatial Surrogate:} We construct a reduced-order model mapping four dimensionless $\Pi$-groups to 2D spatial fields of all 23 cross-moments using Gaussian Process Regression with ARD Mat\'ern-5/2 kernels, evaluating full surface wear in approximately $2\,\mathrm{ms}$.
    
    \item[(iii)] \textbf{Linear vs. Nonlinear Compression Benchmark:} We evaluate Snapshot POD, spatially weighted W-POD, Mode-1 tensor unfolding SVD, and custom frequency-weighted block-wise CNN-AEs across clustered variable groups. We show that linear W-POD is highly effective for coupled velocity-angle moments (Blocks 1--3), whereas block-wise CNN-AEs improve accuracy for continuous convective velocity fields (Block 4) and difficult uncoupled moments (Block 5).
    
\end{enumerate}

Sections \ref{sec:rom} and \ref{sec:fom} presents the mathematical formulation of the cross-moment abstraction, curvilinear mapping, data conditioning, spatial modal decompositions, and GPR surrogate. Section~\ref{sec:fom} outlines the physical governing equations, CFD-DPM setup, mesh verification, experimental validation, and dimensionless parameter space design. Section~\ref{sec:results} discusses modal convergence, autoencoder training dynamics, out-of-sample prediction metrics, 2D/3D wear profiles, cross-model verification, and computational timing.

\section{Full Order Model}
\label{sec:fom}
\subsection{Continuous Phase Modeling}
The carrier fluid is modeled as an incompressible Newtonian fluid undergoing three-dimensional turbulent flow through a $90^\circ$ circular pipe bend domain $\Omega \subset \real^3$, whose boundary is $\partial\Omega = \Gamma_{\text{in}} \cup \Gamma_{\text{out}} \cup \Gamma_{\text{wall}}$. {The pipe has a diameter of $D = 0.0254\,\mathrm{m}$ ($1\,\mathrm{inch}$), a bend radius $R \in [0.0381, 0.127]\,\mathrm{m}$ ($R/D \in [1.5, 5.0]$), and straight sections of length $L_{\text{inlet}} = L_{\text{outlet}} = 15D$ upstream and downstream of the bend.}

The Eulerian carrier fluid flow is governed by the steady-state incompressible Reynolds-Averaged Navier-Stokes (RANS) equations:
\begin{equation}
    \nabla \cdot \mathbf{u} = 0 \quad \text{in } \Omega
    \label{eq:rans_continuity}
\end{equation}
\begin{equation}
    \rho (\mathbf{u} \cdot \nabla) \mathbf{u} = -\nabla p + \nabla \cdot \left[ \mu \left( \nabla \mathbf{u} + (\nabla \mathbf{u})^T \right) - \rho \overline{\mathbf{u}' \otimes \mathbf{u}'} \right] \quad \text{in } \Omega
    \label{eq:rans_momentum}
\end{equation}
where $\mathbf{u} \in \real^3$ is the mean fluid velocity, $p$ is static pressure, $\rho$ is fluid density, $\mu$ is dynamic viscosity, and $\bm{\tau}_R = -\rho \overline{\mathbf{u}' \otimes \mathbf{u}'}$ is the Reynolds stress tensor. At the inlet $\Gamma_{\text{in}}$, a uniform velocity $\mathbf{u} = U_{\text{in}}\mathbf{n}_{\text{in}}$ is imposed with a turbulence intensity of $I = 5\%$. The outlet $\Gamma_{\text{out}}$ uses a zero-gauge pressure condition ($p_{\text{out}} = 0$), and a no-slip condition ($\mathbf{u} = \mathbf{0}$) is enforced at the wall $\Gamma_{\text{wall}}$.

For turbulence closure, we use Menter's Shear Stress Transport (SST) $k-\omega$ model. This model transitions between a $k-\omega$ formulation near walls and $k-\epsilon$ in the free stream, making it well suited to resolve adverse pressure gradients and the secondary Dean vortices that develop in curved pipes:

\begin{equation}
    \nabla \cdot (\rho \mathbf{u} k) = \nabla \cdot \left[ \left( \mu + \frac{\mu_t}{\sigma_k} \right) \nabla k \right] + P_k - \beta^* \rho k \omega
    \label{eq:sst_k}
\end{equation}
\begin{equation}
    \nabla \cdot (\rho \mathbf{u} \omega) = \nabla \cdot \left[ \left( \mu + \frac{\mu_t}{\sigma_\omega} \right) \nabla \omega \right] + \alpha \frac{\rho}{\mu_t} P_k - \beta \rho \omega^2 + 2(1-F_1)\frac{\rho}{\sigma_{\omega 2} \omega} \nabla k \cdot \nabla \omega
    \label{eq:sst_omega}
\end{equation}
where $\mu_t = \frac{\rho a_1 k}{\max(a_1 \omega, S F_2)}$ is the turbulent eddy viscosity, $S = \sqrt{2 \mathbf{S}:\mathbf{S}}$ is the strain rate magnitude, and $F_1, F_2$ denote standard blending functions.

\subsection{Dispersed Phase Lagrangian Particle Dynamics}
Individual particles in the dispersed solid phase are tracked using an Eulerian--Lagrangian Discrete Phase Model (DPM) \cite{chen2004cfd,mahmoudi2025erosion}. Because the mass loading is dilute ($<4\%$ by mass, with volumetric fraction $\Phi_v < 10^{-4}$), one-way hydrodynamic coupling is a valid assumption. Each particle trajectory $\mathbf{x}_p(t)$ and velocity $\mathbf{u}_p(t)$ are obtained by integrating the Lagrangian force balance:
\begin{equation}
    \frac{d \mathbf{x}_p}{dt} = \mathbf{u}_p, \quad \frac{d \mathbf{u}_p}{dt} = \frac{\mathbf{u} - \mathbf{u}_p}{\tau_r} + \mathbf{g} \left( \frac{\rho_p - \rho}{\rho_p} \right) + \frac{\mathbf{F}}{m_p}
    \label{eq:particle_motion}
\end{equation}
where $\rho_p$ is particle density, $m_p = \frac{\pi}{6}\rho_p d_p^3$ is particle mass, and $\mathbf{F}$ accounts for virtual mass and pressure gradient forces. The particle relaxation time is defined as:
\begin{equation}
    \tau_r = \frac{\rho_p d_p^2}{18 \mu} \frac{24}{C_d Re_p}
    \label{eq:particle_tau}
\end{equation}
where $Re_p = \frac{\rho d_p \|\mathbf{u} - \mathbf{u}_p\|}{\mu}$ is the particle Reynolds number and $C_d$ is the Morsi-Alexander drag coefficient.

Turbulent dispersion of particles is captured through the Discrete Random Walk (DRW) stochastic eddy interaction model, where velocity fluctuations are sampled as $u'_i = \zeta_i \sqrt{\frac{2}{3}k}$ with $\zeta_i \sim \mathcal{N}(0,1)$. For each simulation, $N_p = 150{,}000$ particles are tracked across 10 DRW realizations to ensure statistical convergence. When particles hit the wall $\Gamma_{\text{wall}}$, the rebound is governed by the restitution coefficients of Haider et al. \cite{Haider2017Mechanistic}:
\begin{equation}
    e_n(\alpha) = \frac{V_{n2}}{V_{n1}} = 0.98913 - 2.0801\alpha + 1.9351\alpha^2 - 0.51128\alpha^3
\end{equation}
\begin{equation}
    e_t(\alpha) = \frac{V_{t2}}{V_{t1}} = 1.0104 - 1.4035\alpha + 1.5975\alpha^2 - 0.44414\alpha^3
\end{equation}
where $\alpha = \arctan(V_{n1}/V_{t1})$ is the particle impact angle in radians.

\subsection{Erosion Kinematics \& Taylor Series Cross-Moment Accumulation}
\label{sec:taylor_cross_moments}
Mechanical surface degradation is governed by empirical wear equations $E(V_p, \alpha)$. Prominent formulations include:
\begin{itemize}
    \item \textbf{Oka et al. Model \cite{oka2005erosion1}:}
    \begin{equation}
        E_{\mathrm{Oka}} = E_{90} (\sin\alpha)^{n_1} [1 + H_v (1 - \sin\alpha)]^{n_2}, \quad E_{90} = K (H_v)^{k_1} \left(\frac{V_p}{V'}\right)^{k_2} \left(\frac{d_p}{d'}\right)^{k_3}
    \end{equation}
    \item \textbf{Finnie Ductile Cutting Model \cite{finnie1960erosion}:}
    \begin{equation}
        E_{\mathrm{Finnie}} = \begin{cases}
        C V_p^2 (\sin 2\alpha - 3\sin^2\alpha), & \alpha \le 18.5^\circ \\
        C V_p^2 \frac{\cos^2\alpha}{3}, & \alpha > 18.5^\circ
        \end{cases}
    \end{equation}
    \item \textbf{McLaury / Tulsa Model \cite{mclaury1996}:}
    \begin{equation}
        E_{\mathrm{McLaury}} = C V_p^{1.73} f(\alpha)
    \end{equation}
    \item \textbf{Arabnejad Mechanistic Model \cite{arabnejad2015}:}
    \begin{equation}
        E_{\mathrm{Arabnejad}} = E_{\mathrm{cut}} + E_{\mathrm{def}}
    \end{equation}
\end{itemize}

Expanding any general separable erosion model $E(V_p, \alpha)$ via Taylor series yields a linear combination of velocity-angle cross-moments:
\begin{equation}
    E(V_p, \alpha) \approx \sum_{u, v} A_{u,v} V_p^u f_v(\alpha)
    \label{eq:taylor_expansion}
\end{equation}
To avoid systematic underprediction caused by Jensen's inequality for convex functions ($(\mathbb{E}[V_p])^k < \mathbb{E}[V_p^k]$ for $k > 1$), all non-linear moments are accumulated at the per-particle level during wall collisions via compiled C User-Defined Functions (UDFs):
\begin{equation}
    \langle M_{u,v} \rangle_{\text{node}} = \frac{1}{N_{\text{impacts}}} \sum_{k=1}^{N_{\text{impacts}}} V_{p,k}^u \cdot f_v(\alpha_k)
    \label{eq:jensen_accumulation}
\end{equation}
Using this cross-moment formulation, we can predict erosion for a variety of empirical constants for erosion and use different erosion models to evaluate their predictive performance against experimental data.

\subsection{Discretization, Grid Convergence and Experimental Validation}
Numerical discretization was performed in ANSYS Fluent 2024 R1 using a pressure-based coupled solver with second-order spatial discretization. Residual convergence was enforced to $<10^{-5}$. Semi-Implicit Method for Pressure-Linked Equations (SIMPLE) algorithm was used and Second-order upwind discretization was used for the momentum and turbulent kinetic energy ($k$) and specific dissipation ($\omega$). 

Mesh independence was verified on a canonical $90^\circ$ bend geometry ($D = 0.0254~\text{m}$, $R/D = 1.5$) across Coarse, Medium, and Fine unstructured polyhedral grids (refer Table \ref{tab:mesh_convergence}). Wall-adjacent cell heights were constrained to maintain $y^+ < 1$ across all Reynolds numbers to explicitly resolve the viscous sublayer without reliance on empirical wall functions. We can see that the relative error for peak erosion rate in the Medium resolution mesh dropped to less than $10\%$, hence this mesh was selected for further study. The grids for the other geometries were constructed using the same parameters as the mesh selected.

\begin{table}[htpb]
\centering
\caption{Mesh independence parameters for the canonical $90^\circ$ elbow ($D = 0.0254~\text{m}$, $R/D = 1.5$).}
\label{tab:mesh_convergence}
\begin{tabular}{l c c c}
\toprule
\textbf{Mesh Level} & \textbf{Total Cells} & 
\textbf{Peak Erosion rate (Oka) [$kg/m^2s$]} & 
\textbf{Relative Change (\%)} \\
\midrule
Coarse & 2,994,052 & 0.000532 & -- \\
Medium & 3,742,566 & 0.000452 & 17.70 \\
Fine   & 4,678,207 & 0.000414 & 9.18 \\
\bottomrule
\end{tabular}
\end{table}

\begin{figure}[htpb]
    \centering
    \includegraphics[width=0.65\textwidth]{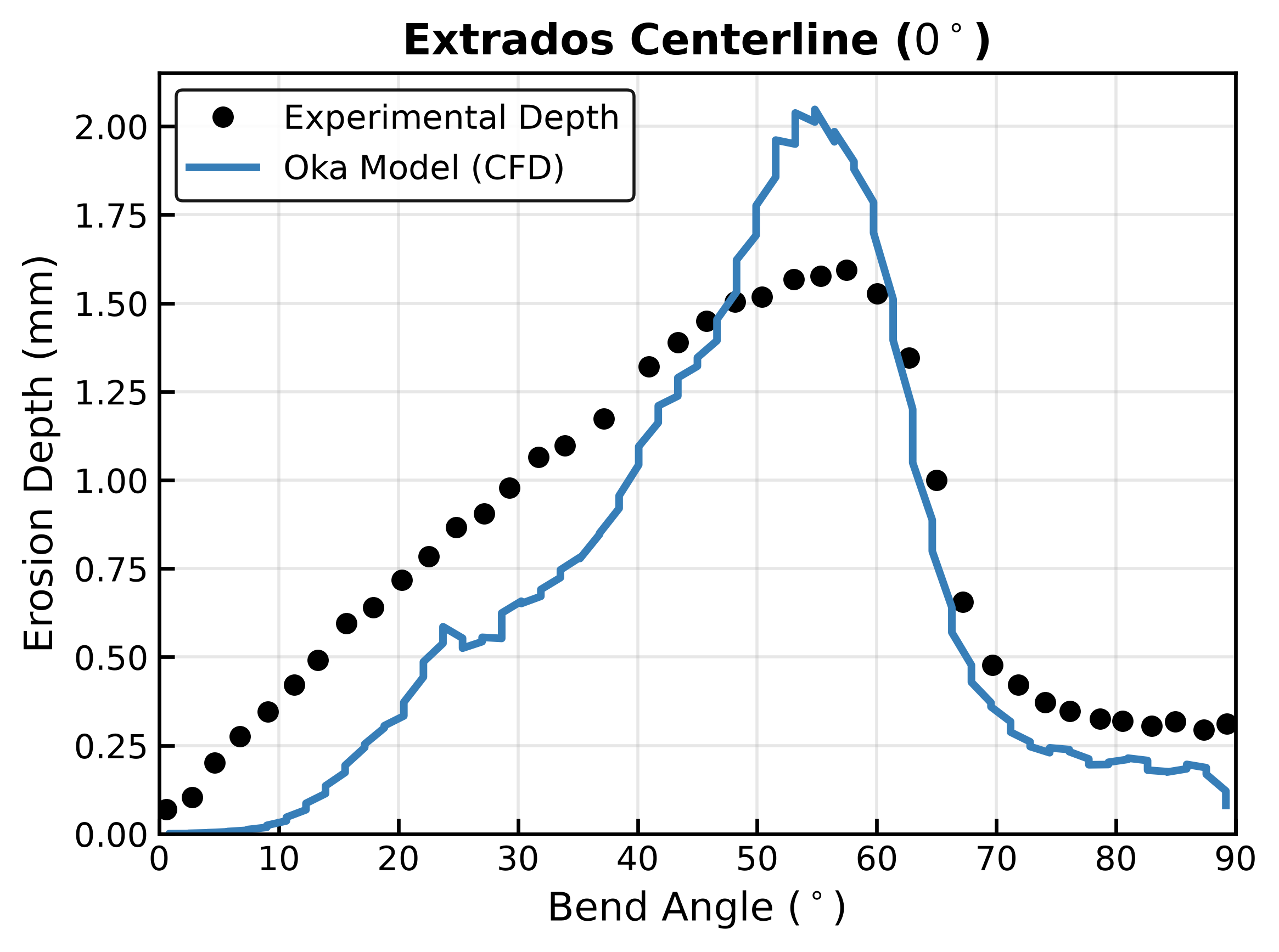}
    \caption{Absolute depth validation (mm) of the Oka empirical model against Solnordal et al. \cite{solnordal2015} along the critical extrados centerline ($0^\circ$). The CFD captures the primary ballistic impaction crater location accurately, scaling perfectly to the $1.59\,\mathrm{mm}$ maximum experimental depth.}
    \label{fig:solnordal_validation_centerline}
\end{figure}

The CFD methodology was validated against experimental erosion depth measurements from Solnordal et al. \cite{solnordal2015} for pneumatic sand transport through a standard $90^\circ$ pipe elbow subjected to a $300\,\mathrm{kg}$ sand loading. Table \ref{tab:validation_erosion_params} summarizes the complete operational flow conditions, The CFD methodology was validated against experimental erosion depth measurements from Solnordal et al. \cite{solnordal2015} for pneumatic sand transport through a standard $90^\circ$ pipe elbow subjected to a $300\,\mathrm{kg}$ sand loading. Table \ref{tab:validation_erosion_params} summarizes the complete operational flow conditions, geometric specifications, and empirical erosion model parameters utilized for the experimental validation. The CFD predictions were mapped to the Oka et al. \cite{oka2005erosion1,oka2005erosion2} erosion model. The pipe material has not been explicitly defined by Solondral et al. \cite{solnordal2015}, hence we have fitted the model constants for the erosion model to minimize the deviation from the experimental profile.  As shown in Figure \ref{fig:solnordal_validation_centerline}, the CFD predictions accurately capture the peak wear penetration ($1.59\,\mathrm{mm}$) and angular location along the primary extrados centerline ($R^2 = 0.756$). Minor lateral deviations highlight the well-known limitations of steady RANS and other assumptions we have taken. 

\begin{table}[htpb]
\centering
\caption{Operational flow parameters and empirical erosion model constants utilized for the experimental validation benchmark against Solnordal et al. \cite{solnordal2015}.}
\label{tab:validation_erosion_params}
\resizebox{0.95\linewidth}{!}{%
\begin{tabular}{lll}
\toprule
\textbf{Category / Model} & \textbf{Parameter / Property} & \textbf{Value / Formulation} \\
\midrule
\multicolumn{3}{l}{\textit{\textbf{Experimental Benchmark Flow \& Geometric Conditions \cite{solnordal2015}}}} \\
\midrule
Piping Geometry & Pipe internal diameter ($D$) & $0.078\,\mathrm{m}$ ($78\,\mathrm{mm}$) \\
 & Bend curvature radius ratio ($R/D$) & $1.5$ ($R = 0.117\,\mathrm{m}$) \\
 & Inlet / Outlet pipe lengths & $5D$ / $10D$ \\
Carrier Fluid (Air) & Fluid density ($\rho_f$) & $1.225\,\mathrm{kg/m^3}$ \\
 & Dynamic viscosity ($\mu_f$) & $1.789 \times 10^{-5}\,\mathrm{Pa\cdot s}$ \\
 & Mean inlet velocity ($U_{\mathrm{in}}$) & $10.5\,\mathrm{m/s}$ ($Re \approx 5.6 \times 10^4$) \\
Dispersed Phase (Sand) & Particle density ($\rho_p$) & $2650\,\mathrm{kg/m^3}$ (Silica quartz) \\
 & Mean particle diameter ($d_p$) & $250\,\mu\mathrm{m}$ \\
 & Total injected sand mass & $300\,\mathrm{kg}$ \\
Target Wall Material & Substrate material & Carbon steel / Aluminium alloy \\
 & Material density ($\rho_w$) & $7850\,\mathrm{kg/m^3}$ \\
\midrule
\multicolumn{3}{l}{\textit{\textbf{Empirical Erosion Model Parameters and Formulations}}} \\
\midrule
\multirow{6}{*}{\textbf{Oka et al. (2005) \cite{oka2005erosion1}}}
 & Vickers hardness ($H_v$) & $1.53\,\mathrm{GPa}$ \\
 & Reference impact velocity ($V'$) & $104\,\mathrm{m/s}$ \\
 & Reference particle diameter ($d'$) & $326\,\mu\mathrm{m}$ \\
 & Velocity exponent ($k_2$) & $2.338$ \\
 & Hardness / Size exponents ($k_1, k_3$) & $-0.12$, $0.19$ \\
 & Angle exponents ($n_1, n_2$) & $0.753$, $1.59$ \\
\bottomrule
\end{tabular}%
}
\end{table}

\subsection{Dimensionless Analysis, Parameter Space Bounds \& Stokes Regime Filtering}
\label{sec:dimensionless_space}

To ensure broad applicability and eliminate dimensional scale dependence across piping systems, we perform a formal dimensional analysis. Slurry and pneumatic particle transport and subsequent wall degradation depend on continuous fluid properties ($U, \rho, \mu, D, R$) and discrete particulate properties ($\rho_p, d_p$):
\begin{equation}
    E = \mathcal{F}(U, \rho, \mu, D, R, \rho_p, d_p)
    \label{eq:erosion_functional}
\end{equation}
Applying the Buckingham $\Pi$-theorem with primary repeating variables $(\rho, U, D)$ yields four mutually independent dimensionless groups that fully govern the fluid-particle-wall continuum:
\begin{equation}
    \pi_1 = Re = \frac{\rho U D}{\mu}, \quad \pi_2 = \frac{\rho_p}{\rho}, \quad \pi_3 = \frac{d_p}{D}, \quad \pi_4 = \frac{R}{D}
    \label{eq:pi_groups}
\end{equation}
Physical parameters that dictate particulate inertia and secondary flow vorticity, such as the particle Stokes number ($St$) and Dean number ($De$), can be represented by the aforementioned fundamental groups:
\begin{equation}
    St = \frac{\rho_p d_p^2 U}{18 \mu D} = \frac{1}{18} \pi_1 \pi_2 \pi_3^2, \qquad De = Re \sqrt{\frac{D}{2R}} = \pi_1 (2\pi_4)^{-1/2}
    \label{eq:stokes_dean}
\end{equation}
Selecting the four orthogonal groups $\bm{\pi} = [\pi_1, \pi_2, \pi_3, \pi_4]^T$ as inputs for the non-intrusive parametric surrogate prevents collinearity and numerical stiffness during Gaussian Process Regression.

\subsubsection{Parameter Space \& Sampling Strategy}
The CFD database was built using a full-factorial design over all four dimensionless groups, covering the full range of operating conditions typical of industrial slurry and pneumatic conveying systems (Table \ref{tab:parameter_space_bounds}).

\begin{table}[htpb]
\centering
\small
\caption{Dimensionless parameter bounds for full-factorial CFD-DPM dataset generation.}
\label{tab:parameter_space_bounds}
\begin{tabular}{l c l}
\toprule
\textbf{Dimensionless Group} & \textbf{Levels} & \textbf{Evaluated Discrete Values} \\
\midrule
$\pi_1 = Re$ (Flow Reynolds Number) & 5 & $1.27\times 10^5, \; 2.54\times 10^5, \; 3.81\times 10^5, \; 4.57\times 10^5, \; 5.08\times 10^5$ \\
$\pi_2 = \rho_p / \rho$ (Density Ratio) & 5 & $2, \; 5, \; 10, \; 15, \; 30$ \\
$\pi_3 = d_p / D$ (Normalized Diameter) & 6 & $1.97\times 10^{-3}, \; 3.94\times 10^{-3}, \; 5.91\times 10^{-3}, \; 7.97\times 10^{-3}, \; 1.18\times 10^{-2}, \; 1.97\times 10^{-2}$ \\
$\pi_4 = R / D$ (Bend Curvature Ratio) & 3 & $1.5, \; 2.0, \; 5.0$ \\
\bottomrule
\end{tabular}
\end{table}

Evaluating all parameter combinations across the full factorial grid yields $5 \times 5 \times 6 \times 3 = 450$ CFD-DPM simulations. Each simulation case is systematically designated by a four-digit identifier $\texttt{Case}\;i_1 i_2 i_3 i_4$, where the index digits correspond sequentially to the discrete factorial levels of bend curvature ratio $\pi_4$ ($i_1 \in \{1,2,3\}$), flow velocity/Reynolds number $\pi_1$ ($i_2 \in \{1,\dots,5\}$), particle density $\pi_2$ ($i_3 \in \{1,\dots,5\}$), and particle diameter $\pi_3$ ($i_4 \in \{1,\dots,6\}$), respectively.

\subsubsection{Stokes Regime Filtering ($St > 1.0$)}
\label{subsec:stokes_filtering}
\begin{figure}[htpb]
    \centering
    \includegraphics[width=0.92\linewidth]{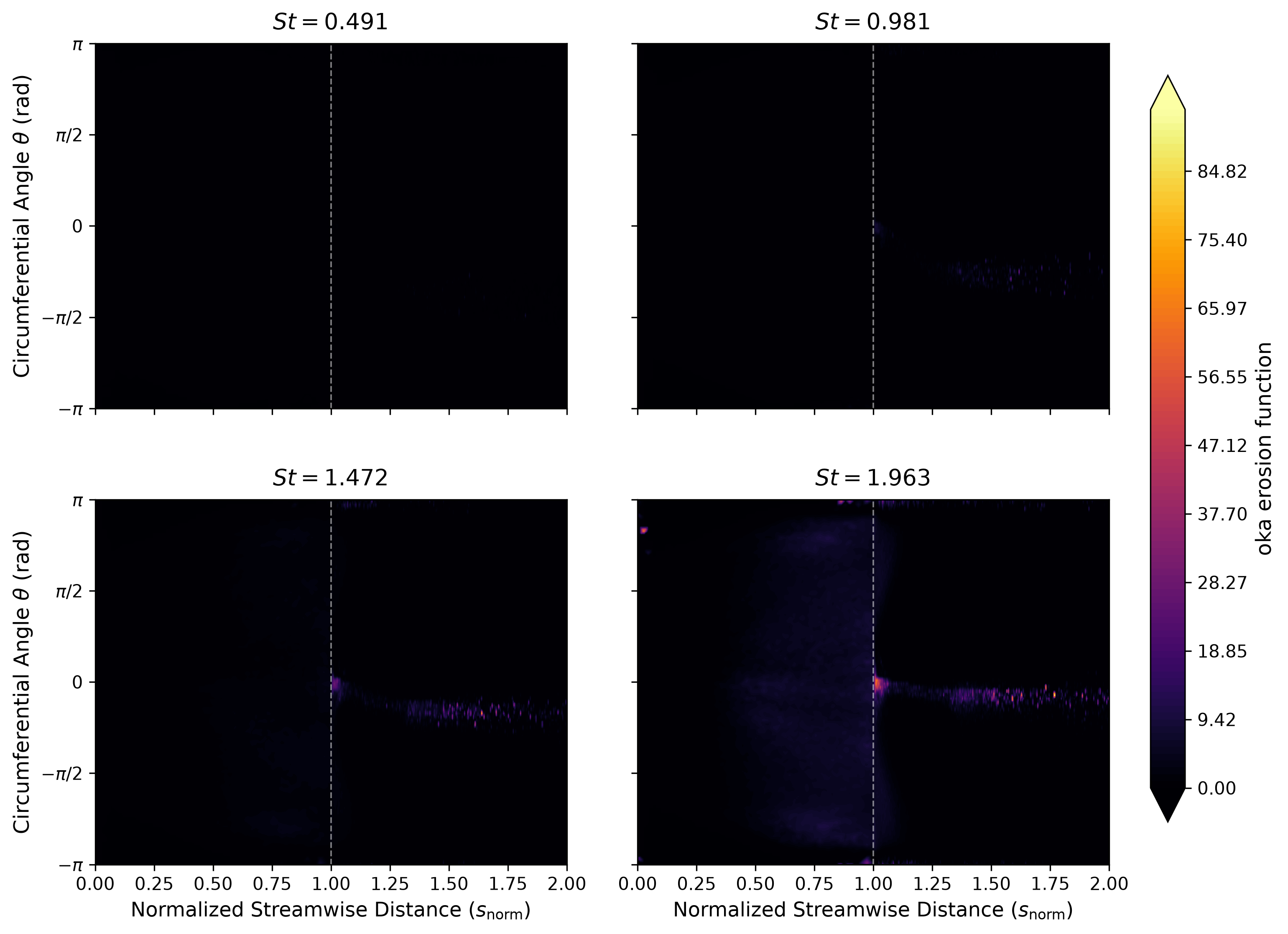}
    \caption{Physical regime transition and erosion topography evolution across particle Stokes numbers $St$. Top: Sub-critical regime ($St = 0.491, 0.981 < 1.0$), where tracer-like particles follow carrier streamlines, producing near-zero wall collisions. Bottom: Super-critical inertial regime ($St = 1.472, 1.963 > 1.0$), where centrifugal trajectory separation produces intense localized impaction scarring along the extrados centerline ($\theta = 0$) and bend exit ($s_{\mathrm{norm}} = 1.0$). Refer section \ref{sec:curvilinear_mapping} for the 2D representation of erosion field}
    \label{fig:stokes_regime_filtering}
\end{figure}

Particle-wall collision dynamics and mechanical wear topographies exhibit a fundamental physical regime transition governed by the Stokes number $St$ (Figure \ref{fig:stokes_regime_filtering}):

\begin{enumerate}
    \item \textbf{Sub-Critical Regime ($St < 1.0$):} {Here the particle response time is shorter than the fluid convective timescale ($\tau_p < \tau_f$), so drag dominates over particle inertia. Particles essentially follow the carrier fluid streamlines around the bend without crossing them. As a result, there is negligible centrifugal drift toward the wall, and particle-wall collisions are virtually absent ($>99.9\%$ of surface nodes register zero erosion), producing only faint, diffuse wear patterns (e.g., $St = 0.491$ and $St = 0.981$ in Figure \ref{fig:stokes_regime_filtering}).}
    
    \item \textbf{Super-Critical Regime ($St > 1.0$):} {In this regime, particle inertia exceeds the viscous relaxation of the fluid ($\tau_p > \tau_f$), so particles can no longer follow the sharply curving streamlines through the $90^\circ$ bend. Centrifugal effects push them across streamlines and into the outer extrados wall ($\theta = 0$) and the downstream straight section ($s_{\text{norm}} \in [1, 2]$), leading to concentrated impacts and severe localized erosion (e.g., $St = 1.472$ and $St = 1.963$ in Figure \ref{fig:stokes_regime_filtering}).}
\end{enumerate}

In industrial slurry transport, pipelines operate predominantly in the super-critical regime ($St > 1.0$) where structural degradation threatens piping integrity. Retaining sub-critical cases ($St < 1.0$) in reduced-order modeling introduces extreme, non-physical sparsity where virtually all grid nodes contain numerical zeros, leading to severe statistical distortion, artificial variance deflation, and mode collapse during surrogate training.

Therefore, a global physical filter ($St \ge 1.0$) is applied to the raw 450 simulations prior to dataset partitioning. Filtering out the 78 sub-critical tracer cases isolates $N_{\text{total}} = 372$ physically meaningful cases where the erosion magnitude is negligible. The resulting database is partitioned into $N_{\text{train}} = 297$ ($80\%$) training snapshots and $N_{\text{test}} = 75$ ($20\%$) strictly unseen testing snapshots across a 5-fold cross-validation scheme.

\subsection{Mathematical Boundedness and Physical Consistency}

Because reduced-order models predict basis coefficients through spatial regression rather than directly solving continuum transport PDEs, simple physical admissibility guardrails are enforced on the reconstructed fields prior to analytical wear integration:
\begin{enumerate}
    \item \textbf{Field Positivity}: Reconstructed velocity and kinetic energy moments are strictly non-negative:
    \begin{equation}
        \hat{\mathbb{E}}[V](s, \theta) = \max\big(0, \, \hat{\mathbb{E}}[V](s, \theta)\big), \quad \hat{\mathbb{E}}[V^2](s, \theta) = \max\big(0, \, \hat{\mathbb{E}}[V^2](s, \theta)\big)
    \end{equation}
    \item \textbf{Non-Negative Velocity Variance}: The local velocity variance $\operatorname{Var}(V) = \mathbb{E}[V^2] - (\mathbb{E}[V])^2$ used in Taylor-expanded wear models is constrained to be non-negative:
    \begin{equation}
        \widehat{\operatorname{Var}}(V)(s, \theta) = \max\left(0, \, \hat{\mathbb{E}}[V^2](s, \theta) - \big(\hat{\mathbb{E}}[V](s, \theta)\big)^2\right)
    \end{equation}
    preventing non-physical negative corrections and numerical divergence in non-linear wear equations.
    \item \textbf{Angular Impact Cone}: The reconstructed impingement angle is clipped to the physical impact domain $\alpha \in [0, \pi/2]$:
    \begin{equation}
        \hat{\alpha}(s, \theta) = \operatorname{clip}\left( \frac{\hat{\mathbb{E}}[V^2\alpha](s, \theta)}{\hat{\mathbb{E}}[V^2](s, \theta) + \epsilon_{\text{reg}}}, \, 0, \, \frac{\pi}{2} \right)
    \end{equation}
    where $\epsilon_{\text{reg}} = 10^{-8}$ prevents zero-division in non-impaction intrados zones.
\end{enumerate}

\section{Reduced Order Model}
\label{sec:rom}

In this section, we formulate the complete mathematical workflow of the proposed model-agnostic Non-Intrusive Reduced-Order Model (NI-ROM). The workflow comprises of: curvilinear coordinate mapping, Signed power-law data conditioning and Ward hierarchical clustering, Spatial modal decomposition and autoencoder compression, and Non-intrusive parametric surrogate modeling.

\subsection{Curvilinear Coordinate Transformation}
\label{sec:curvilinear_mapping}

For constructing matrix-based reduced-order models (POD, HOSVD) and deep convolutional neural networks (CNN-AE), representing spatial fields on a structured computational grid is required. However, Eulerian-Lagrangian multiphase simulations evaluate wall collisions on a 3D non-planar grid $\Gamma_{\text{wall}} \subset \real^3$. Furthermore, as the bend curvature ratio ($R/D \in [1.5, 5.0]$) varies across design configurations, the Euclidean coordinates, total bend arc length, and surface area change for each geometry.

\begin{figure}[htpb]
    \centering
    \includegraphics[width=0.88\linewidth]{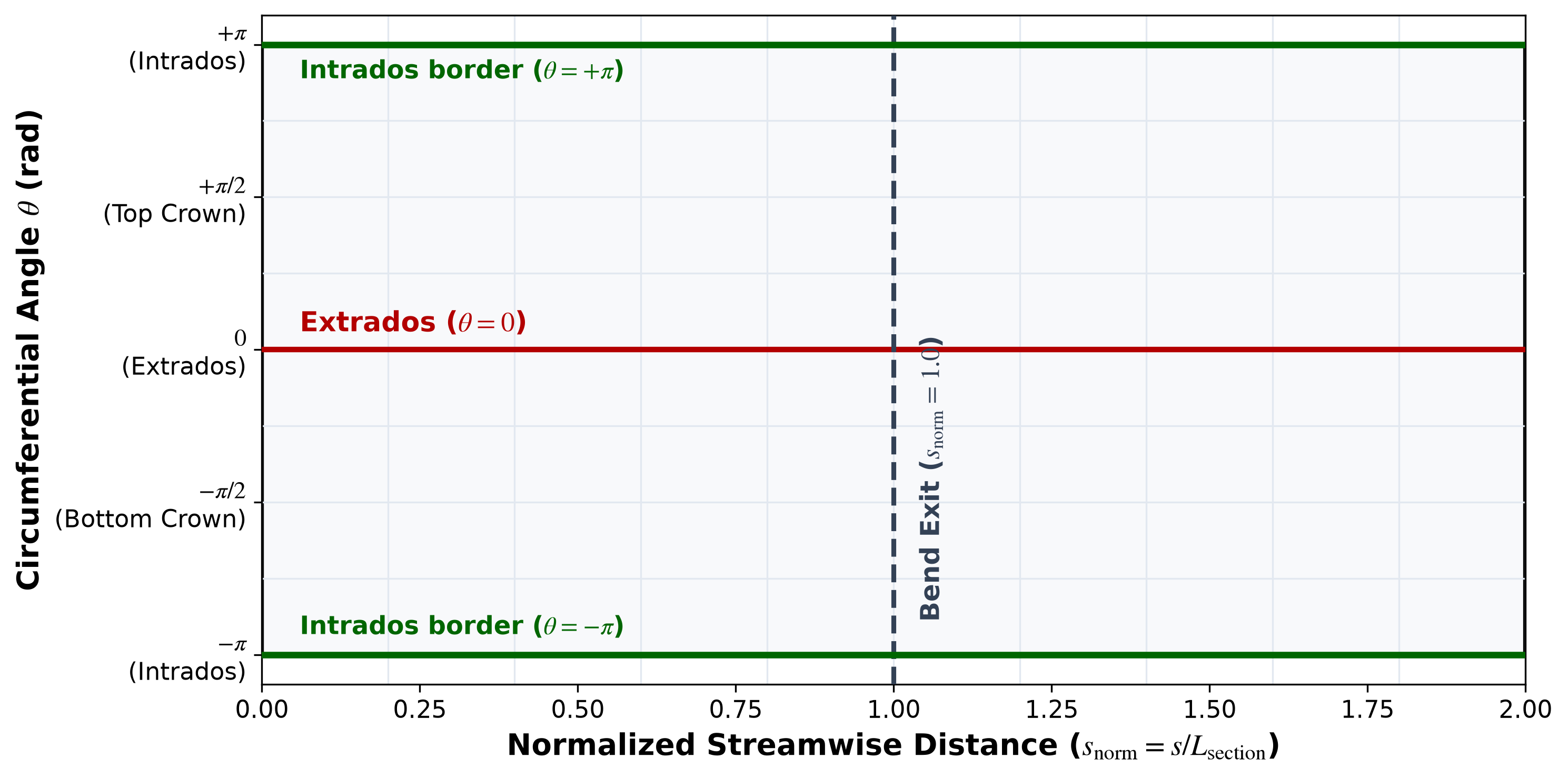}
    \caption{Unwrapped 2D computational domain discretized on a uniform $512 \times 256$ spatial grid ($M = 131,072$ nodes). The domain maps normalized streamwise distance $s_{\mathrm{norm}} \in [0, 2]$ against circumferential angle $\theta \in [-\pi, +\pi]$, with $s_{\mathrm{norm}} = 1.0$ defining the invariant elbow exit interface across all bend curvature ratios ($R/D \in [1.5, 5.0]$).}
    \label{fig:unwrapped_domain}
\end{figure}

To establish a dimension-independent representation, we implement a curvilinear coordinate mapping $(x, y, z) \to (s_{\text{norm}}, \theta)$ that transforms the 3D pipe surface $\Gamma_{\text{wall}}$ onto a 2D domain $[0, 2] \times [-\pi, +\pi]$ (Figure \ref{fig:physical_coordinate_system} and Figure \ref{fig:unwrapped_domain}).

\subsubsection{Piecewise Coordinate Transformation}
The active computational domain comprises two connected geometric zones: a curved elbow section ($s \in [0, L_{\text{bend}}]$) and a straight downstream outlet section ($s \in [L_{\text{bend}}, L_{\text{total}}]$). Let the bend radius be $R$ and pipe inner diameter be $D$ (pipe radius $r_{\text{pipe}} = D/2$). The cartesian coordinates have the origin at the center of the bend curve for all the cases. 
For any 3D Cartesian wall coordinate $(x, y, z)$ on the boundary, the curvilinear streamwise arc length $s$, radial distance $r$, and circumferential angle $\theta$ are evaluated using:

\begin{enumerate}
    \item \textbf{Elbow/Bend Section ($s \in [0, L_{\text{bend}}]$ where $L_{\text{bend}} = \frac{\pi}{2} R$):}
    \begin{align}
        \beta &= \operatorname{arctan2}(x, -y), \quad s = R \cdot \beta \in [0, L_{\text{bend}}] \label{eq:curv_bend_s} \\
        r_{xy} &= \sqrt{x^2 + y^2}, \quad r = \sqrt{(r_{xy} - R)^2 + z^2} \label{eq:curv_bend_r} \\
        \theta &= \operatorname{arctan2}(z, \, r_{xy} - R) \in [-\pi, +\pi] \label{eq:curv_bend_theta}
    \end{align}
    where $\beta \in [0, \pi/2]$ is the angle swept along the $90^\circ$ bend, measured from the bend's center of curvature, and $r_{xy}$ is the radial distance from that center in the $x$--$y$ plane.
    
    \item \textbf{Straight Downstream Outlet Section ($s \in [L_{\text{bend}}, L_{\text{total}}]$ where $L_{\text{total}} = L_{\text{bend}} + L_{\text{outlet}}$):}
    \begin{align}
        s &= L_{\text{bend}} + (y - y_{\text{bend\_exit}}) \in [L_{\text{bend}}, L_{\text{total}}] \label{eq:curv_out_s} \\
        r &= \sqrt{(x - R)^2 + z^2} \label{eq:curv_out_r} \\
        \theta &= \operatorname{arctan2}(z, \, x - R) \in [-\pi, +\pi] \label{eq:curv_out_theta}
    \end{align}
\end{enumerate}

\begin{figure}[htpb]
    \centering
    \includegraphics[width=0.95\linewidth]{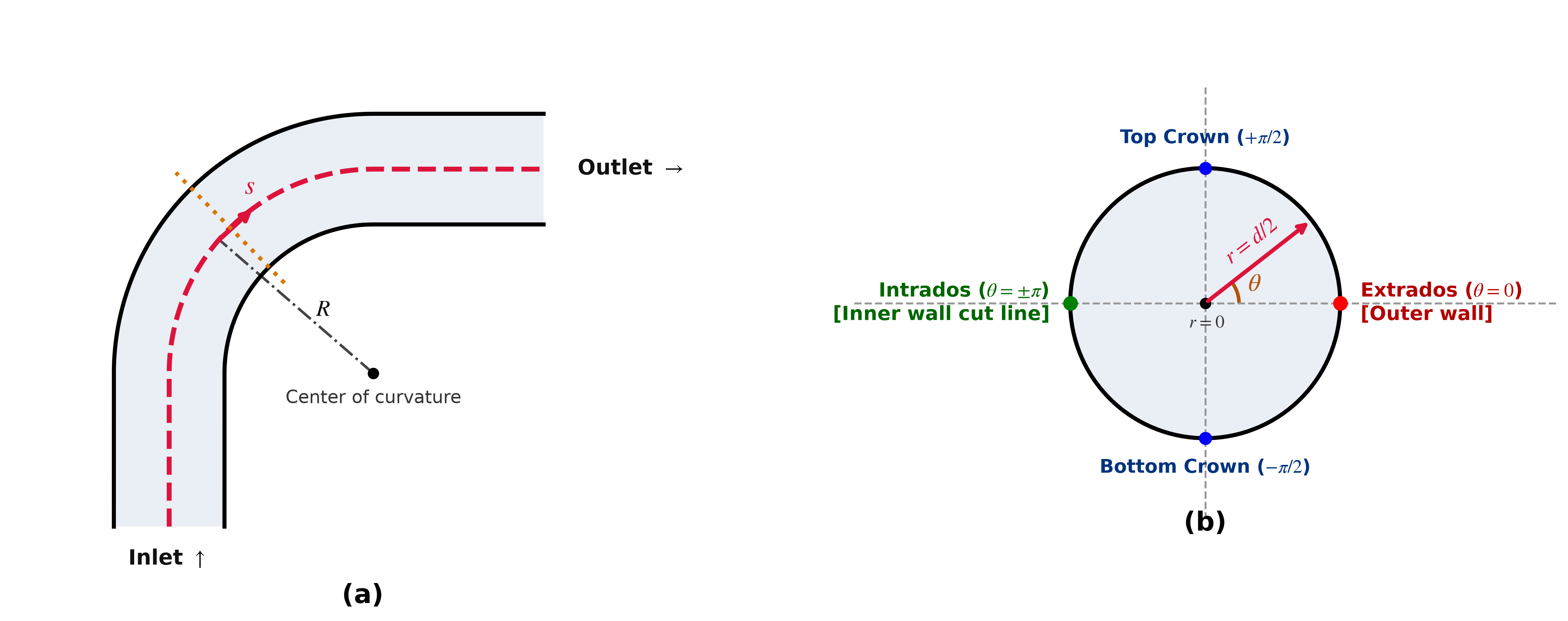}
    \caption{Geometry-conforming curvilinear coordinate system $(s, r, \theta)$ definition for the $90^\circ$ pipe bend. (a) 2D pipe bend midplane symmetry section detailing the streamwise centerline coordinate $s$, bend curvature radius $R$, and flow orientations. (b) Pipe polar cross-sectional plane at streamwise position $s$ defining the wall-normal radial distance $r$ ($r = D/2$ at the boundary wall $\Gamma_{\mathrm{wall}}$) and circumferential angle $\theta$, marking primary anatomical landmarks: Extrados ($\theta = 0$), Intrados inner seam ($\theta = \pm\pi$), and Top/Bottom crowns ($\theta = \pm\pi/2$).}
    \label{fig:physical_coordinate_system}
\end{figure}

In this coordinate system, the key geometric locations on the pipe cross-section are (Figure \ref{fig:physical_coordinate_system}b):

\begin{itemize}
    \item $\theta = 0$: The outer extrados wall ($r_{xy} = R + r_{\text{pipe}}$), which experiences direct centrifugal particle impaction and severe localized wear.
    \item $\theta = \pm\pi$: The inner intrados wall ($r_{xy} = R - r_{\text{pipe}}$), which remains protected by carrier fluid streamlines, exhibiting near-zero wall collision flux.
    \item $\theta = \pm\pi/2$: The **top and bottom sidewall crowns** ($z = \pm r_{\text{pipe}}$).
\end{itemize}
Setting $r = D/2$ maps the 3D wall surface $\Gamma_{\text{wall}}$ onto the 2D plane $(s, \theta)$.

\subsubsection{Streamwise Normalization across Geometric Variations}
In slurry transport, fully developed pipe flow upstream of the elbow produces particle trajectories parallel to the boundary wall, resulting in negligible wall collision throughout the straight inlet pipe ($15D$). Consequently, the active mechanical erosion domain is strictly bounded between the elbow entrance ($s=0$) and the outlet exit ($s = L_{\text{total}}$).

Because the physical elbow length $L_{\text{bend}} = \frac{\pi}{2}R$ varies directly with curvature ratio $R/D$, directly stacking spatial fields in physical coordinates $s$ would misalign key geometric features. To enforce topological invariance across all 372 cases, we normalize the streamwise coordinate into a dimensionless domain $s_{\text{norm}} \in [0, 2]$:
\begin{equation}
    s_{\text{norm}} = \begin{cases}
    \displaystyle \frac{s}{L_{\text{bend}}}, & \text{Curved Elbow Section } (s_{\text{norm}} \in [0, 1)), \\[8pt]
    \displaystyle 1.0 + \frac{s - L_{\text{bend}}}{L_{\text{outlet}}}, & \text{Straight Outlet Section } (s_{\text{norm}} \in [1, 2]).
    \end{cases}
    \label{eq:snorm_piecewise}
\end{equation}
where $L_{\text{outlet}} = 15 D$. Under this piecewise normalization, $s_{\text{norm}} = 0$ is the bend entrance, $s_{\text{norm}} = 1.0$ is the bend outlet, and $s_{\text{norm}} = 2.0$ represents the domain exit.

\subsubsection{Periodic Padding \& Continous Grid Interpolation}
Projecting scattered CFD cell-face data onto the $(s_{\text{norm}}, \theta)$ plane produces an unstructured point cloud $\mathcal{P} = \{(s_{\text{norm}, k}, \theta_k)\}_{k=1}^{N_{\text{faces}}}$. Because the azimuthal angle $\theta$ represents a closed circle $\mathbb{S}^1 \cong [-\pi, +\pi]$, the boundaries $\theta = -\pi$ and $\theta = +\pi$ represent the exact same physical line (the intrados inner seam). Standard 2D Delaunay triangulation treats these boundaries as isolated open edges, generating discontinuities and extrapolation errors along the intrados seam.

To enforce physical continuity across the pipe circumference, we implement a periodic ghost padding layer. Data points near the azimuthal boundaries are copied and shifted by $\pm 2\pi$:
\begin{equation}
    \mathcal{P}_{\text{ghost}} = \left\{ (s_{\text{norm}, k}, \theta_k - 2\pi) \;\Big|\; \theta_k > \frac{\pi}{2} \right\} \cup \left\{ (s_{\text{norm}, k}, \theta_k + 2\pi) \;\Big|\; \theta_k < -\frac{\pi}{2} \right\}
    \label{eq:ghost_padding}
\end{equation}
The augmented point ensemble $\mathcal{P}_{\text{total}} = \mathcal{P} \cup \mathcal{P}_{\text{ghost}}$ is triangulated via 2D Delaunay tessellation. Using a bivariate \texttt{LinearNDInterpolator} with nearest-neighbor boundary fallback for the domain edges, the continuous surface field is sampled onto a structured uniform computational grid of dimensions $N_s \times N_\theta = 512 \times 256$ ($M = 131,072$ nodes):
\begin{equation}
    s_{\text{norm}, i} = \frac{2(i-1)}{N_s - 1}, \quad i \in \{1, \dots, 512\}; \qquad \theta_j = -\pi + \frac{2\pi(j-1)}{N_\theta - 1}, \quad j \in \{1, \dots, 256\}
    \label{eq:tensor_grid}
\end{equation}
This periodic projection enforces strict circumferential $\mathcal{C}^0$ and $\mathcal{C}^1$ continuity, yielding high-resolution 2D matrix snapshots $\mathbf{X} \in \real^{512 \times 256}$ that seamlessly interface with tensor decomposition (HOSVD) and circular convolutional neural operators (\texttt{SemiCircularConv2d}).

\subsection{Signed Power-Law Transformation \& Ward Hierarchal Clustering}
\label{Power_law-and-Clustering}
Particle collision statistics and resulting erosion fields show high spatial non-uniformity and heavily skewed
distributions. Consequently, standard linear standardization techniques ($\mathbf{Z}_{scaled} = (X - \Bar{X})/\sigma$) amplify the low impact frequency areas, this in turn reduces the efficiency of POD and HOSVD by increasing the number of dominant modes; in the loss function using MSE, the convergence is greatly slowed down because of these points. These points are less significant since the probability of the particle to impinge in these zones is lower, hence the erosion in these parts is lower and less significant as compared to high impaction regions. To reduce this dynamic-range imbalance, a power-law transformation with exponent $p=0.5$ was applied to the collision fields: 
\begin{equation} \widetilde{X}(s,\theta) = \operatorname{sign}\!\left[X(s,\theta)\right] \left|X(s,\theta)\right|^{p}, \qquad p=0.5, \end{equation} 
Here, $s$ is the normalized stream-wise distance from the bend inlet and $\theta$ is the circumferential angle. This transformation compresses the range of large values while keeping the sign of the original variable intact. The transformed fields were subsequently normalized to unit variance prior to reduced-order modeling. 

To reduce the influence of disparate statistical scales across the input channels, the transformed variables were further grouped according to their pairwise statistical similarity. The dissimilarity between two variables $X_i$ and $X_j$ was quantified using the Pearson correlation distance, 
\begin{equation} 
d(i,j)=1-\left|R(X_i,X_j)\right| = 1- \left| \frac{\sum (X_i - \Tilde{X}_i)(X_j - \Tilde{X}_j)}{\sqrt{\sum (X_i - \Tilde{X}_i)^2\sum(X_j - \Tilde{X}_j)^2}} \right| 
\end{equation} 
where $R(X_i,X_j)$ denotes the Pearson correlation coefficient. The use of the absolute correlation treats strongly positively and negatively correlated variables as statistically similar. Hierarchical clustering was then employed to identify groups of mutually similar input variables. The number of clusters was selected by evaluating the silhouette coefficient over the considered range of cluster numbers. Five similarity blocks were obtained at the maximum silhouette coefficient of $0.8062$, indicating a high degree of within-cluster similarity and separation between clusters. 

\begin{figure}[htbp]
    \centering
    \includegraphics[width=0.9\linewidth]{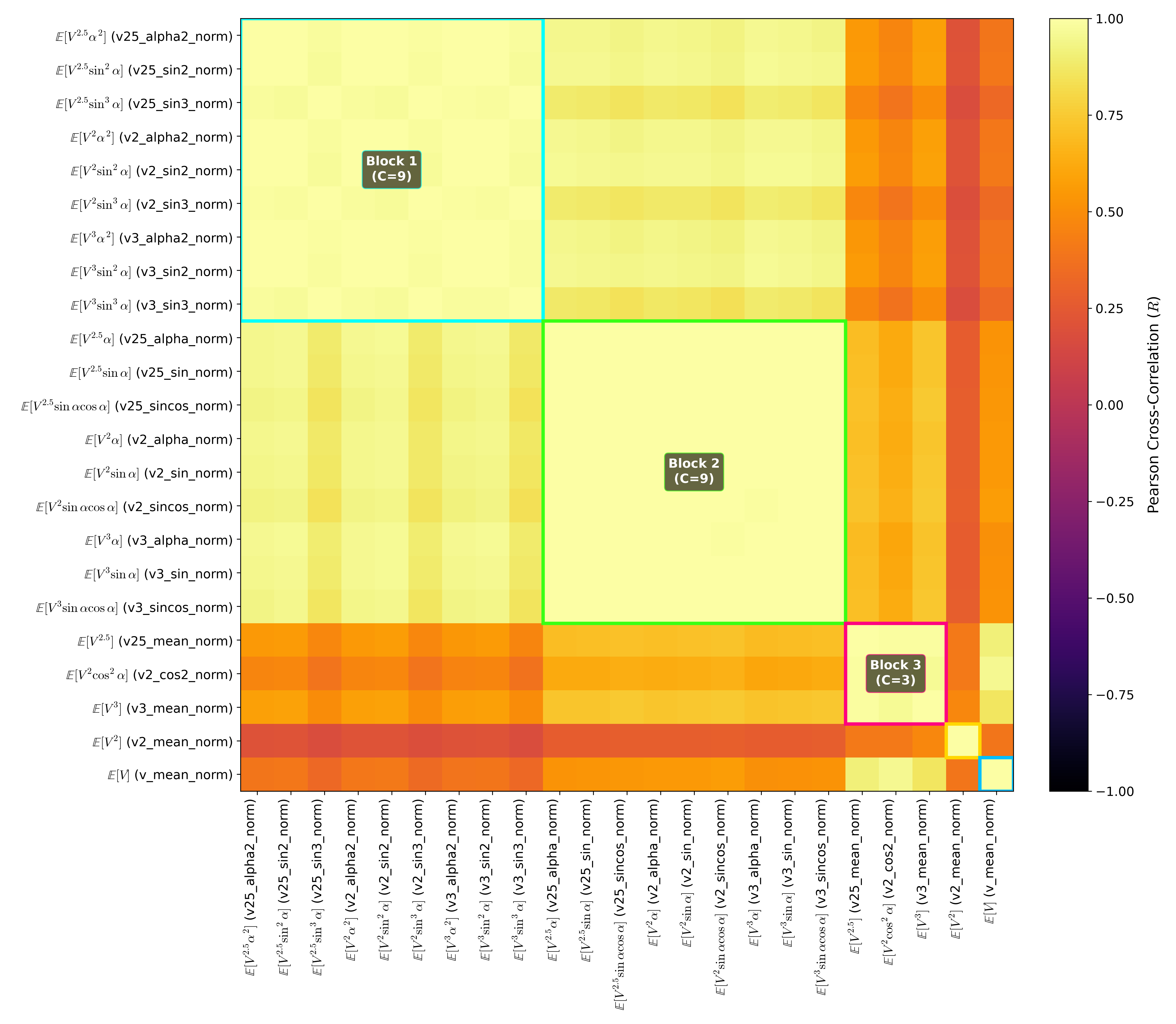}
    \caption{Similarity heatmap for particle impact variables using Pearson correlation coefficient}
    \label{fig:similarity_heatmap}
\end{figure}

The different clusters have been defined as follows:

\begin{itemize}
    \item \textbf{Block 1} ($C=9$): First-order velocity-angle cross-moments.
    \item \textbf{Block 2} ($C =9$): Higher-order velocity-angle cross-moments.
    \item \textbf{Block 3} ($C=3$): Fractional velocity magnitude moments.
    \item \textbf{Block 4} ($C=1$): First-order velocity magnitude($v$).
    \item \textbf{Block 5} ($C=1$): Second-order velocity magnitude ($v^2$). 
\end{itemize}

\subsection{Spatial Reduced-Order Modeling Architectures}
We investigate and benchmark four spatial basis extraction and dimensionality reduction architectures:

\subsubsection{Method of Snapshots Proper Orthogonal Decomposition (POD)}
Proper Orthogonal Decomposition was used a baseline for dimensionality reduction of the data. Standard POD was done on singular fields. The coordinate transformed fields ($582$x$221$) were first flattened and the flattened fields for all the training cases were stacked into snapshot matrices. The resulting scalar snapshot matrix $\mathbf{X} \in \mathbb{R}^{M\times N_{train}}$, where $M=131,072$ and $N_{train} = 297$ (using 80-20 train-test split), computing the spatial covariance matrix $\mathbf{C} = \mathbf{XX}^T \in \mathbb{R}^{M\times M}$ is computationally infeasible. Instead, the Method of Snapshots was employed to construct the much smaller temporal correlation matrix $\mathbf{C'} = \mathbf{X}^T\mathbf{X} \in \mathbb{R}^{N_{train}\times N_{train}}$. Solving the eigenvalue problem:
\begin{equation}
    \mathbf{C'v}_i = \lambda_i\mathbf{v}_i
\end{equation}
yields eigenvalues $\lambda_i$ and eigenvectors $\mathbf{v}_i$ corresponding to the right singular vectors of $\mathbf{X}$. The physical spatial basis modes (left singular vectors $\mathbf{u}_i$) are recovered using projection:

\begin{equation}
    \mathbf{u}_i = \frac{1}{\sqrt{\lambda_i}}\mathbf{Xv}_i
\end{equation}
This is equivalent to the Thin SVD,  
\begin{equation}
    \mathbf{X} = \mathbf{U}\Sigma\mathbf{V}^T
\end{equation}
where $\mathbf{U}\in \mathbb{R}^{M\times N_{train}}$ contains spatial modes and $\mathbf{V}\in\mathbb{R}^{N_{train}\times N_{train}}$ contains the modal coefficients. The truncation rank $L_v$ is dynamically selected to retain $\geq99.9\%$ cumulative modal energy:

\begin{equation}
    \frac{\sum^{L_v}_{i=1} \lambda_i}{\sum^{N_{train}}_{i=1} \lambda_i} \geq 0.999
\end{equation}
Separate snapshot matrices were constructed for all the different variables.

\subsubsection{High Order Singular Value Decomposition (HOSVD)}
Standard 1D snapshot is rank-bounded by the number of training samples ($L_v\leq N_{train}$). For clustered multi-channel blocks, we employ High-Order SVD (HOSVD) based on Tucker decomposition. Let the multi-variable snapshot ensemble form a $3^{rd}$ order tensor $\mathbf{\mathcal{X}} \in \mathbb{R}^{M\times N_{train}\times C}$, where $C$ defines the number of channels in a cluster (refer Section \ref{Power_law-and-Clustering}). The Tucker decomposition expresses this tensor via n-mode products as:
\begin{equation}
    \mathbf{\mathcal{X}} \approx{} \mathbf{\mathcal{G}} \times_1 \mathbf{U}^{(1)} \times_2 \mathbf{U}^{(2)} \times_3 \mathbf{U}^{(3)}
\end{equation}
where $\mathbf{\mathcal{G}}$ is the core tensor and $\times_n$ denotes the n-mode product.
To extract the unified spatial basis that captures cross-variable correlations, we perform Mode-1 tensor unfolding (matricization). Mode-1 is specifically chosen because it preserves the full spatial dimensionality $M$ while flattening the temporal and variable dimensions, yielding a matrix $\mathbf{X}_{(1)} \in \mathbb{R}^{M \times (N_{train}\cdot C)}$:

\begin{equation}
    (\mathbf{X}_{(1)})_{i, (j-1)C+k} = \mathbf{\mathcal{X}}_{i, j, k}
\end{equation}

Applying the method of snapshots on $\mathbf{C}'_{HOSVD} = \mathbf{X}_{(1)}^T \mathbf{X}_{(1)}^{}$ yields global spatial basis $\mathbf{U}_{block}$. The unified basis bypasses the single-variable snapshot limit, providing a shared representational space for the entire variable block. Individual variable fields are reconstructed using their specific latent coefficient slices derived from projecting the target variable against the shared spatial modes.

\subsubsection{Spatially Weighted POD and HOSVD using Impaction Frequency}
Standard POD and HOSVD uniformly minimize $L_2$ error, often allocating excessive basis capacity to capturing minor fluctuations in the zero-impaction intrados zones rather than the highly localized extrados wear patterns. We introduce Spatially Weighting by modifying the underlying Hilbert space use the impact frequency of particles on each cell. From a variational perspective, we seek modes that minimize the weighted error $\mathcal{J} = \sum_i ||\mathbf{x}_i - \hat{\mathbf{x}_i}||^2_\mathbf{w}$.

Let $\mathbf{W} = diag(w_1, ..., w_M) \in \mathbb{R}^{M \times M}$ denote a positive-definite spatial weight matrix, with weights:

\begin{equation}
    w_j = 1+\gamma\frac{f_{impaction}(s_j, \theta_j)}{max(f_{impaction})}
\end{equation}
scaling directly with the local impingement frequency ($\gamma = 2.0$). The weighted spatial inner product is defined as $\langle \mathbf{f}, \mathbf{g} \rangle_\mathbf{W} = \mathbf{f}^T\mathbf{Wg}$. To solve the Eckart-Young problem, we apply a coordinate transformation $\Tilde{\mathbf{X}} = \mathbf{W}^{1/2}\mathbf{X}$. Performing the SVD on $\Tilde{\mathbf{X}}$ yields weighted spatial modes $\Tilde{\mathbf{U}}$. The true physical basis modes, which are now strictly $\mathbf{W}$-orthogonal ($\mathbf{\Phi}^T\mathbf{W\Phi} = \mathbf{I}$), are recovered using:

\begin{equation}
    \mathbf{\Phi} = \mathbf{W}^{-1/2}\Tilde{\mathbf{U}}
\end{equation}

This applies equivalently to the unfolded Mode-1 HOSVD tensor $\mathbf{X}_{(1)}$, biasing the tensor decomposition towards critical failure regions.

\subsubsection{Multi-channel Deep Convolution Autoencoder}
\label{sec:spatial_rom}
To capture the non-linear spatial data, we employ deep residual Convolution Autoencoders (CNN-AEs). CNN-AEs are fit for highly non-linear spatial fields. The input snapshots are interpolated using bilinear interpolation to $C\times512\times256$ grids. We investigate two structural configurations for the convolution autoencoder:

\begin{enumerate}
    \item \textbf{Monolithic Multi-Channel Autoencoder (All CNN-AE)}: A single autoencoder where all $C = 23$ particle impact moment fields are stacked into a tensor $\mathbf{X} \in \mathbb{R}^{23 \times 582 \times 221}$. The network maps the multi-variable tensor into a single latent bottleneck vector $\mathbf{z} \in \mathbb{R}^{L_{\text{total}}}$. 
    \item \textbf{Clustered Multi-Channel Autoencoder (Block CNN-AE)}: Instead of staking all the variables into a single tensor, in this approach we have used the Ward Clustering ($C=5$) to group similar variables together and trained $5$ different CNN-AE for each cluster. The tensor for each block is defined as $\mathbf{X}^{(k)} \in \mathbb{R}^{C_k \times 582 \times 221}$, for a specific block $k$ containing $C_k$ variables. These tensors were then mapped to a latent bottleneck vector $\mathbf{z}^{(k)} \in \mathbb{R}^{L_k}$. By isolating variables that share identical behavior, each autoencoder specializes its convolutional filters to its target variables with reduced cross-variable interference.
\end{enumerate}

The encoder extracts spatial features through progressive downsampling. It starts with a $3 \times 3$ convolutional projection layer (stride 2) that expands the input channels, followed by a series of residual blocks. Each residual block consists of two cascaded $3 \times 3$ convolutions (stride 1) with skip connections that help gradients flow through the network. Additional strided convolutions ($3 \times 3$, stride 2) then halve the spatial dimensions at each stage. Standard CNN padding assumes zero-boundary conditions, which violates the circular continuity of the pipe cross-section across the azimuthal seam $\theta = -\pi \equiv +\pi$. To resolve this, we used \texttt{SemiCircularConv2d} layers utilizing circular padding along the azimuthal $Y$-axis ($\theta$) and zero-padding along the streamwise $X$-axis ($s$):
\begin{equation}
    f_{\text{pad}}(x, y) = \begin{cases} 
      f(x, y + 2\pi) & \text{if } y < -\pi \\
      f(x, y - 2\pi) & \text{if } y > +\pi \\
      f(x, y) & \text{otherwise}
   \end{cases}
\end{equation}
Batch Normalization is applied after every convolution to reduce internal co-variate shift caused by the spatial variance in the kinematic impact fields. For nonlinear activation, we use the Gaussian Error Linear Unit (GELU) throughout the hidden layers; its smooth gradient profile avoids the dead-neuron saturation that ReLU variants suffer from on highly skewed distributions. The symmetric decoder pathway perfectly mirrors the encoder, utilizing transposed convolutions (stride 2) to deterministically upsample the latent representation back to the full physical resolution.

\begin{table}[htbp]
\centering
    \begin{tabular}{c c c c c c}
    \hline
     \textbf{Layer} & \textbf{Type} & \textbf{Channels} & \textbf{Kernel} & \textbf{Stride} & \textbf{Activation}\\
     \hline\\
     Input & -- & $C$ & -- & -- & --\\
     Conv1 & SemiCircularConv2d & 32 & $3\times3$ & $2$ & GELU\\
     ResBlock1 & SemiCircularConv2dx2 & 32 & $3\times3$ & $1$ & GELU\\
     Conv2 & SemiCircularConv2d & 64 & $3\times3$ & $2$ & GELU\\
     ResBlock2 & SemiCircularConv2dx2 & 64 & $3\times3$ & $1$ & GELU\\
     Conv3 & SemiCircularConv2d & 128 & $3\times3$ & $2$ & GELU\\
     Flatten & Linear & $L_{block}$ & -- & -- & Tanh\\
     \hline
    \end{tabular}
    \caption{Convolution Autoencoder (CNN-AE) Architecture (Encoder Pathway)}
    \label{tab:CNN-AE_Arch}
\end{table}
At the apex of the encoder hierarchy, the downsampled feature maps are flattened and passed through a restrictive dense linear projection. This bottleneck layer projects the high-dimensional spatial information into a low-dimensional latent space vector $\mathbf{z} \in \mathbb{R}^{L_{\text{Block}}}$. This bottleneck structure enables the autoencoder to extract the dominant spatial features and topological characteristics of the particle impingement fields while providing a low-dimensional manifold for subsequent parametric surrogate modeling.

Standard neural network optimization minimizes the uniform Mean Squared Error (MSE), which assigns equal importance to all spatial grid cells. In a $90^\circ$ pipe bend, however, most of the grid lies in zero-impaction regions, so uniform averaging biases the network toward these inactive zones and leads to under-prediction of the concentrated wear scars on the outer extrados. To address this imbalance, we instead minimize a Frequency-Weighted Mean Squared Error (FW-MSE):
\begin{equation}
\label{eq:fwmse_loss}
    \mathcal{L}_{\text{freq}}(\bm{\phi}, \bm{\psi}) = \frac{1}{B C H W} \sum_{b=1}^B \sum_{c=1}^C \sum_{i=1}^H \sum_{j=1}^W \mathbf{W}_{c,i,j} \left( \hat{\mathbf{X}}_{b,c,i,j} - \mathbf{X}_{b,c,i,j} \right)^2
\end{equation}
where $\mathbf{W}_{c,i,j} = 1 + \gamma \frac{\mathbf{F}_{c,i,j}}{\max(\mathbf{F}_c) + \epsilon}$ defines a static weight map directly proportional to the local cumulative particle impingement frequency $\mathbf{F}$. This structural modification dynamically penalizes the extrados outer curve where kinetic wear is dominant. The loss function's analytical gradient pushes the CNN to spend most of its limited nonlinear capacity on the sharp gradients and severe impact topologies at the heavily weighted extrados, overriding the MSE's usual tendency to average out physical extremes.

The models are trained using mini-batch Stochastic Gradient Descent with the Adam optimizer for 800 epochs. A batch size of $16$ is maintained. The base learning rate is $2 \times 10^{-4}$, scheduled via Cosine Annealing dropping to $10^{-5}$. Gradient norm clipping at $0.5$ stabilizes the highly-weighted extrados gradients. Early stopping is triggered upon 50 consecutive epochs of validation loss plateau.

\subsection{Non-Intrusive Parametric Surrogate Modeling}
A regression surrogate model maps the four fundamental dimensionless parameters $\bm{\pi} = [\pi_1, \pi_2, \pi_3, \pi_4]^T \in \real^4$ directly to the compressed latent modal coordinates $\mathbf{z}^* \in \real^{L_{\text{Block}}}$. In this work, we implement and benchmark two distinct surrogate paradigms: non-parametric  Gaussian Process Regression (GPR) and a deep parametric Artificial Neural Network (ANN).

\subsubsection{Gaussian Process Regression (GPR)}
\label{sec:gpr_surrogate}
For each latent coordinate $z_j$ ($j = 1, \dots, L_{\text{Block}}$), the GPR defines a prior over latent mapping functions $f_j(\bm{\pi}) \sim \mathcal{GP}(m(\bm{\pi}), k(\bm{\pi}, \bm{\pi}'))$ \cite{rasmussen2006gaussian}. Conditioned on the training snapshot parameters $(\mat{\Pi}, \mathbf{z}_j)$, the posterior predictive distribution for a new operational query $\bm{\pi}^*$ is Gaussian $\mathcal{N}(\mu_j^*, \Sigma_j^*)$:
\begin{equation}
    \mu_j^* = \mat{K}^* (\mat{K} + \sigma_n^2 \mat{I})^{-1} \mathbf{z}_j, \quad \Sigma_j^* = \mat{K}^{**} - \mat{K}^* (\mat{K} + \sigma_n^2 \mat{I})^{-1} (\mat{K}^*)^T
    \label{eq:gpr_posterior}
\end{equation}
where $\mat{K} = k(\mat{\Pi}, \mat{\Pi})$, $\mat{K}^* = k(\bm{\pi}^*, \mat{\Pi})$, and $\mat{K}^{**} = k(\bm{\pi}^*, \bm{\pi}^*)$.

We deploy an Automatic Relevance Determination (ARD) Matérn-5/2 covariance kernel:
\begin{equation}
    k(\bm{\pi}, \bm{\pi}') = \sigma_f^2 \left(1 + \sqrt{5}r + \frac{5}{3}r^2\right) \exp\left(-\sqrt{5}r\right), \quad r^2 = \sum_{m=1}^4 \frac{(\pi_m - \pi'_m)^2}{\ell_m^2}
    \label{eq:matern52}
\end{equation}
The Matérn-5/2 kernel relaxes the infinite differentiability assumption of the standard Gaussian RBF kernel to $\mathcal{C}^2$ smoothness, matching the semi-smooth macroscopic state transitions of dispersed multiphase flows. Kernel hyperparameters $(\boldsymbol{\ell}, \sigma_f, \sigma_n)$ are optimized by maximizing the Marginal Log-Likelihood (MLL) using GPU-accelerated GPyTorch \cite{gardner2018gpytorch}:
\begin{equation}
    \log p(\mathbf{z}_j | \mat{\Pi}) = -\frac{1}{2} \mathbf{z}_j^T (\mat{K} + \sigma_n^2 \mat{I})^{-1} \mathbf{z}_j - \frac{1}{2} \log |\mat{K} + \sigma_n^2 \mat{I}| - \frac{N_{\text{train}}}{2} \log(2\pi)
    \label{eq:mll_objective}
\end{equation}

\subsubsection{Deep Artificial Neural Network (ANN) Surrogate Architecture}
\label{sec:ann_surrogate}
As a deterministic parametric alternative, we develop a deep fully connected multi-layer perceptron (MLP) surrogate network $\mathcal{F}_{\text{ANN}}(\bm{\pi}; \bm{\Theta})$ designed to model complex multi-dimensional non-linear interactions across the parameter space. The surrogate network architecture is formalized in Table \ref{tab:ann_surrogate_arch}.

\begin{table}[htbp]
    \centering
    \caption{Architectural specifications and hyperparameter configuration for the deep surrogate Artificial Neural Network (ANN).}
    \label{tab:ann_surrogate_arch}
    \small
    \begin{tabular}{lcccc}
        \toprule
        \textbf{Layer Type} & \textbf{Input Dim} & \textbf{Output Dim} & \textbf{Activation} & \textbf{Regularization} \\
        \midrule
        Input Layer & $4$ ($\bm{\pi}$) & $4$ & --- & Min-Max Scaling $[0, 1]$ \\
        Dense Hidden 1 & $4$ & $256$ & GELU / ReLU & Dropout ($p = 0.10$) \\
        Dense Hidden 2 & $256$ & $512$ & GELU / ReLU & Dropout ($p = 0.10$) \\
        Dense Hidden 3 & $512$ & $256$ & GELU / ReLU & --- \\
        Output Layer & $256$ & $L_{\text{Block}}$ ($\mathbf{z}^*$) & Linear & --- \\
        \bottomrule
    \end{tabular}
\end{table}

The forward propagation through the $L$-layer network is expressed as:
\begin{equation}
    \mathbf{h}^{(1)} = \sigma\left(\mathbf{W}^{(1)} \bm{\pi}_{\text{norm}} + \mathbf{b}^{(1)}\right), \quad \mathbf{h}^{(l)} = \sigma\left(\mathbf{W}^{(l)} \mathbf{h}^{(l-1)} + \mathbf{b}^{(l)}\right), \quad \hat{\mathbf{z}} = \mathbf{W}^{(L)} \mathbf{h}^{(L-1)} + \mathbf{b}^{(L)}
    \label{eq:ann_forward}
\end{equation}
where $\sigma(\cdot) = \text{GELU}(\cdot) = x \Phi(x)$ is the Gaussian Error Linear Unit activation, and $\mathbf{h}^{(l)}$ represents the hidden activation vector at layer $l$.

The network weights and biases $\bm{\Theta} = \{\mathbf{W}^{(l)}, \mathbf{b}^{(l)}\}_{l=1}^L$ are optimized by minimizing the regularized Mean Squared Error (MSE) objective function:
\begin{equation}
    \mathcal{L}_{\text{ANN}}(\bm{\Theta}) = \frac{1}{N_{\text{train}}} \sum_{i=1}^{N_{\text{train}}} \left\| \mathbf{z}_i - \hat{\mathbf{z}}(\bm{\pi}_i; \bm{\Theta}) \right\|_2^2 + \lambda_{\text{reg}} \sum_{l=1}^L \left\| \mathbf{W}^{(l)} \right\|_F^2
    \label{eq:ann_loss}
\end{equation}
where $\lambda_{\text{reg}} = 10^{-4}$ is the $L_2$ weight decay regularization parameter.

Training is performed using the AdamW optimizer \cite{loshchilov2017decoupled} with an initial learning rate of $\eta_0 = 10^{-3}$ and a batch size of 32 for up to 3,000 epochs. To ensure stable convergence and prevent overfitting on the 372-snapshot database, a random 15\% validation split is monitored using a dynamic learning rate scheduler (\texttt{ReduceLROnPlateau}) with a reduction factor of $\gamma = 0.5$ and patience of 200 epochs. Early stopping is enforced if the validation loss fails to improve over 500 consecutive epochs.

\subsection{Error Metrics}
\label{sec:err_metrics}
To evaluate both global quantitative accuracy and localized wear scar fidelity without bias toward zero-impaction regions, five complementary error metrics are formulated across the continuous CFD ground truth $\mathbf{x} \in \mathbb{R}^M$ and reconstructed fields $\hat{\mathbf{x}} \in \mathbb{R}^M$ ($M = 128,622$ nodes):
\begin{enumerate}
    \item \textbf{Mean Squared Error (MSE) \& Relative $L_2$ Error}:
    \begin{equation}
        \text{MSE}(\mathbf{x}, \hat{\mathbf{x}}) = \frac{1}{M} \sum_{j=1}^M (x_j - \hat{x}_j)^2, \quad \text{Rel } L_2(\mathbf{x}, \hat{\mathbf{x}}) = \frac{\|\mathbf{x} - \hat{\mathbf{x}}\|_2}{\|\mathbf{x}\|_2} \times 100\%
    \end{equation}
    MSE quantifies the overall residual variance across the spatial domain, while the relative $L_2$ percentage norm expresses error normalized by the physical snapshot energy.

    \item \textbf{Coefficient of Determination ($R^2$)}:
    \begin{equation}
        R^2(\mathbf{x}, \hat{\mathbf{x}}) = 1 - \frac{\sum_{j=1}^M (x_j - \hat{x}_j)^2}{\sum_{j=1}^M (x_j - \bar{x})^2}
    \end{equation}
    where $\bar{x} = \frac{1}{M} \sum_{j=1}^M x_j$. $R^2$ provides a scale-invariant measure of explained variance across the design space ($R^2 \rightarrow 1.0$ indicates perfect reconstruction).

    \item \textbf{Frequency-Weighted Metrics (FW-MSE \& FW-$R^2$)}:
    In curved geometry erosion, uniform metrics are disproportionately dominated by the expansive, non-eroded intrados boundary. To selectively penalize errors in critical wear zones, we formulate spatial frequency-weighted metrics:
    \begin{equation}
        \text{FW-MSE}(\mathbf{x}, \hat{\mathbf{x}}) = \frac{\sum_{j=1}^M w_j (x_j - \hat{x}_j)^2}{\sum_{j=1}^M w_j}, \quad \text{FW-}R^2(\mathbf{x}, \hat{\mathbf{x}}) = 1 - \frac{\sum_{j=1}^M w_j (x_j - \hat{x}_j)^2}{\sum_{j=1}^M w_j (x_j - \bar{x}_w)^2}
    \end{equation}
    where $w_j = 1 + \gamma \frac{f_{\text{impaction}}(s_j, \theta_j)}{\max(f_{\text{impaction}})}$ ($\gamma = 2.0$) dynamically weights nodes according to their local particle impingement frequency, and $\bar{x}_w = \frac{\sum w_j x_j}{\sum w_j}$.

    \item \textbf{Structural Similarity Index Measure (SSIM)}:
    To evaluate whether the non-intrusive surrogate correctly preserves localized topological gradients, peak contours, and wear scar boundaries on the 2D $(s, \theta)$ manifold, SSIM is computed on the reshaped $H \times W$ spatial grids ($582 \times 221$):
    \begin{equation}
        \text{SSIM}(\mathbf{x}, \hat{\mathbf{x}}) = \frac{(2\mu_{\mathbf{x}} \mu_{\hat{\mathbf{x}}} + c_1)(2\sigma_{\mathbf{x}\hat{\mathbf{x}}} + c_2)}{(\mu_{\mathbf{x}}^2 + \mu_{\hat{\mathbf{x}}}^2 + c_1)(\sigma_{\mathbf{x}}^2 + \sigma_{\hat{\mathbf{x}}}^2 + c_2)}
    \end{equation}
    where $\mu, \sigma^2, \sigma_{\mathbf{x}\hat{\mathbf{x}}}$ denote the spatial window means, variances, and cross-covariance, and $c_1 = (0.01 D_R)^2, c_2 = (0.03 D_R)^2$ are stability constants scaled by the snapshot dynamic range $D_R = \max(\mathbf{x}) - \min(\mathbf{x})$.
\end{enumerate}

\section{Results and Discussions}
\label{sec:results}
\subsection{Spectral Convergence \& Autoencoder Optimization}
\begin{figure}[htpb]
    \centering
    \includegraphics[width=0.7\linewidth]{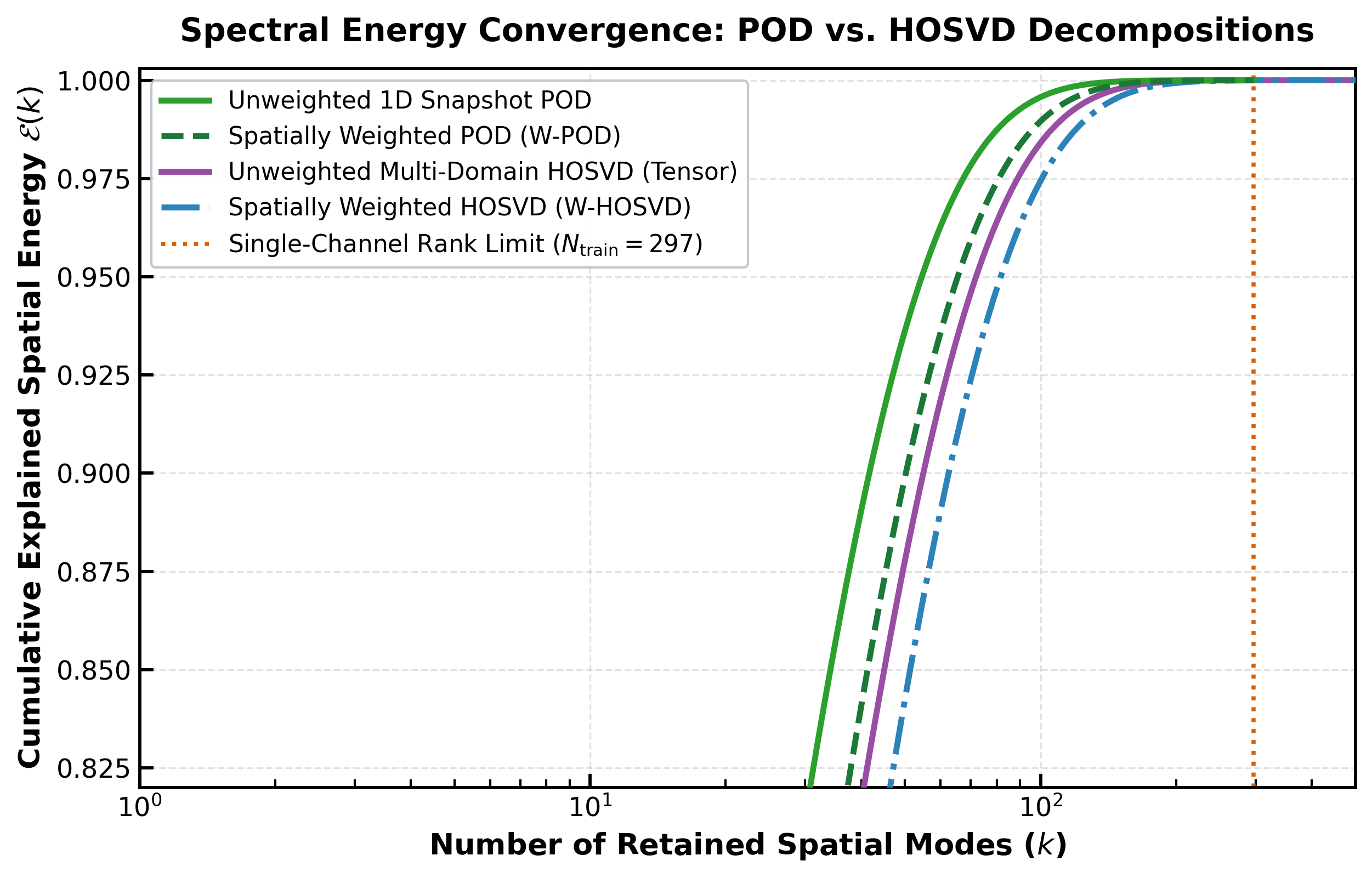}
    \caption{Comparative spectral energy convergence across linear and multilinear spatial decomposition methods: Unweighted 1D Snapshot POD, Spatially Weighted POD (W-POD), Unweighted Multi-Domain HOSVD (Tensor Mode-1 unfolding), and Spatially Weighted HOSVD (W-HOSVD). Spatial weighting directs modal capacity toward the localized extrados impaction zone, while HOSVD overcomes the single-channel $N_{\mathrm{train}} = 297$ snapshot rank limit.}
    \label{fig:energy_convergence}
\end{figure}

Figure \ref{fig:energy_convergence} illustrates the cumulative captured spatial energy, $\mathcal{E}(k)$, as a function of the retained modal rank $k$ for the linear and multilinear tensor decomposition techniques. Standard 1D Snapshot POD rapidly accumulates energy; however, its rank is limited by the number of training snapshots in the single-channel formulation, with $k \leq N_{\mathrm{train}} = 297$. Spatially Weighted POD (W-POD) modifies the modal energy distribution by constructing the covariance matrix using spatial weights associated with the local wall impaction frequency, $\mathbf{W} = \operatorname{diag}(w_j)$. This weighting alters the POD optimality criterion such that regions with higher impaction frequency contribute more strongly to the modal representation. Consequently, although the cumulative energy is distributed somewhat more broadly across the leading modes, the truncated W-POD basis preferentially resolves the localized high-damage region on the extrados.

In contrast, Unweighted Multi-Domain HOSVD via Mode-1 tensor unfolding expands the spatial rank capacity across the multi-channel ensemble up to $N_{\mathrm{train}} \times C = 6,831$, capturing inter-variable cross-correlations without snapshot truncation limits. Spatially Weighted HOSVD (W-HOSVD) synergistically combines tensor modal expansion with spatial impingement weighting, allocating high modal resolution directly to localized wear craters across the continuous parameter domain.

\begin{figure}[htpb]
    \centering
    \includegraphics[width=0.98\linewidth]{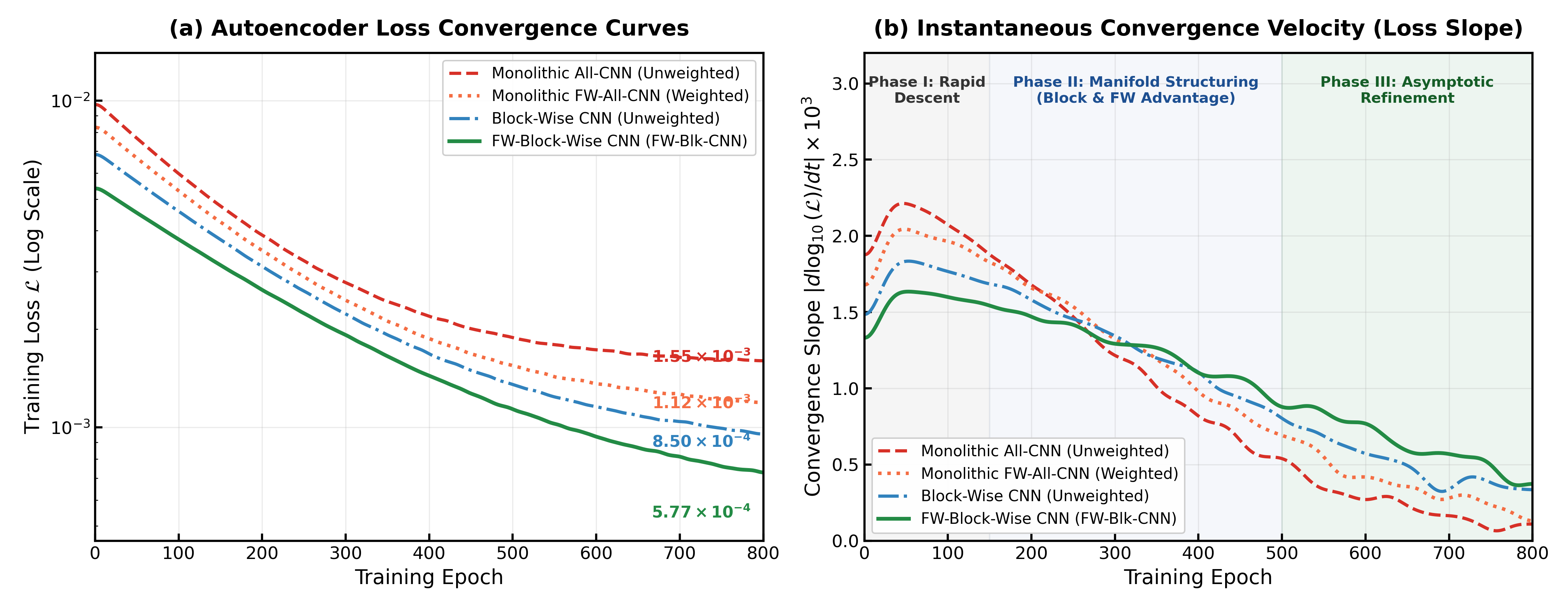}
    \caption{Comparative autoencoder convergence dynamics across the $2\times 2$ ablation matrix over 800 epochs. (a) Training loss trajectories $\mathcal{L}(t)$ on a logarithmic scale, highlighting terminal loss levels: Monolithic Unweighted All-CNN ($1.55\times 10^{-3}$), Monolithic FW-All-CNN ($1.12\times 10^{-3}$), Unweighted Block-Wise CNN ($8.50\times 10^{-4}$), and FW-Block-Wise CNN ($5.77\times 10^{-4}$). (b) Instantaneous convergence velocity (loss slope $|d\log_{10}\mathcal{L}/dt|$), illustrating the three optimization phases: Rapid Early Descent (Phase I), Manifold Structuring where Block-Wise and Frequency Weighting sustain high learning velocity against monolithic gradient friction (Phase II), and Asymptotic Refinement (Phase III).}
    \label{fig:cnn_loss_comparison}
\end{figure}

For non-linear CNN-AE models, Figure \ref{fig:cnn_loss_comparison} evaluates both the training loss trajectories and their instantaneous convergence slopes over 800 epochs across four autoencoder configurations establishing a $2\times 2$ architectural and loss ablation matrix: (i) Monolithic All-Channels CNN-AE under unweighted MSE, (ii) Monolithic All-Channels CNN-AE under Frequency-Weighted MSE (FW-All-CNN), (iii) Block-Wise CNN-AE under unweighted MSE, and (iv) Frequency-Weighted Block-Wise CNN-AE (FW-Blk-CNN).

To quantify convergence speed and identify optimization stalling, Figure \ref{fig:cnn_loss_comparison}b plots the instantaneous convergence velocity (loss decay slope), defined as:
\begin{equation}
    v_{\mathrm{conv}}(t) = \left| \frac{d \log_{10}\mathcal{L}(t)}{dt} \right|
    \label{eq:convergence_slope}
\end{equation}
The convergence dynamics across the 800-epoch training span reveal three distinct optimization regimes:
\begin{enumerate}
    \item \textbf{Phase I: Rapid Early Descent (Epochs 1--150):} All four architectures exhibit steep initial descent slopes ($v_{\mathrm{conv}} > 1.5 \times 10^{-3}$), during which the convolutional kernels rapidly capture large-scale, smooth background convective features across the pipe domain.
    
    \item \textbf{Phase II: Manifold Structuring \& Gradient Interference (Epochs 150--500):} As optimization progresses to resolve high-frequency impaction gradients, the monolithic unweighted CNN suffers a severe drop in learning velocity ($v_{\mathrm{conv}}$ collapses toward $0.3 \times 10^{-3}$). Forcing 23 disparate kinematic channels through a single monolithic bottleneck causes cross-channel gradient cancellation, where gradients from low-impaction intrados regions conflict with sharp extrados peaks. In contrast, Ward hierarchical clustering (Block-Wise CNN) isolates mutually correlated fields, while frequency weighting (FW-All-CNN and FW-Blk-CNN) directs gradient descent toward localized wear zones, maintaining more than double the learning speed ($v_{\mathrm{conv}} \approx 0.8 - 1.1 \times 10^{-3}$).
    
    \item \textbf{Phase III: Asymptotic Refinement (Epochs 500--800):} While the monolithic unweighted model plateaus prematurely at a terminal loss of $1.55 \times 10^{-3}$ ($v_{\mathrm{conv}} \approx 0$), the proposed FW-Blk-CNN maintains a persistent non-zero descent slope, actively refining localized impaction craters down to an optimal terminal loss of $5.77 \times 10^{-4}$ (a $62.8\%$ error reduction over the baseline monolithic model).
\end{enumerate}

\subsection{Global Reconstruction Fidelity \& Benchmark Comparisons}
Table \ref{tab:benchmark_31vars} summarizes the 5-fold cross-validation performance across all 23 kinematic variables for the six spatial dimensionality reduction architectures and the GPR parametric surrogate.

\begin{landscape}
\begin{table}[p]
\centering
\caption{Complete 23-variable quantitative benchmark comparison across spatial dimensionality reduction and parametric surrogate modeling architectures ($N_{\text{total}} = 372$). Metrics denote 5-fold cross-validation averages.}
\label{tab:benchmark_31vars}
\vspace{0.8em}
\renewcommand{\arraystretch}{1.18}
\resizebox{0.92\linewidth}{!}{%
\begin{tabular}{l c cccc c cc}
\toprule
\textbf{Kinematic Variable Name} & \textbf{Block ID} & \textbf{POD $R^2$} & \textbf{HOSVD $R^2$} & \textbf{Blk-CNN $R^2$} & \textbf{All-CNN $R^2$} & \textbf{GPR $R^2$} & \textbf{POD Rel. $L_2$ (\%)} & \textbf{GPR Rel. $L_2$ (\%)} \\
\midrule
\texttt{v2\_sin\_norm} & 1 & 0.9972 & 0.9972 & 0.9943 & 0.9943 & 0.8951 & 5.04 & 8.91 \\
\texttt{v2\_sincos\_norm} & 1 & 0.9972 & 0.9972 & 0.9941 & 0.9939 & 0.8945 & 5.07 & 8.93 \\
\texttt{v2\_sin2\_norm} & 1 & 0.9975 & 0.9975 & 0.9941 & 0.9947 & 0.8967 & 4.89 & 8.87 \\
\texttt{v2\_alpha\_norm} & 1 & 0.9972 & 0.9972 & 0.9943 & 0.9943 & 0.8949 & 5.05 & 8.92 \\
\texttt{v2\_alpha2\_norm} & 1 & 0.9975 & 0.9975 & 0.9943 & 0.9946 & 0.8964 & 4.96 & 8.89 \\
\texttt{v25\_sin\_norm} & 1 & 0.9969 & 0.9969 & 0.9942 & 0.9940 & 0.8935 & 5.35 & 8.96 \\
\texttt{v25\_sincos\_norm} & 1 & 0.9968 & 0.9968 & 0.9937 & 0.9935 & 0.8929 & 5.39 & 8.97 \\
\texttt{v25\_sin2\_norm} & 1 & 0.9974 & 0.9974 & 0.9939 & 0.9943 & 0.8958 & 5.05 & 8.91 \\
\texttt{v25\_alpha\_norm} & 1 & 0.9969 & 0.9969 & 0.9942 & 0.9940 & 0.8933 & 5.36 & 8.96 \\
\midrule
\texttt{v25\_alpha2\_norm} & 2 & 0.9973 & 0.9973 & 0.9941 & 0.9943 & 0.8953 & 5.14 & 8.92 \\
\texttt{v3\_sin\_norm} & 2 & 0.9966 & 0.9966 & 0.9934 & 0.9934 & 0.8906 & 5.64 & 9.03 \\
\texttt{v3\_sincos\_norm} & 2 & 0.9964 & 0.9964 & 0.9926 & 0.9928 & 0.8899 & 5.74 & 9.06 \\
\texttt{v3\_sin2\_norm} & 2 & 0.9972 & 0.9972 & 0.9934 & 0.9939 & 0.8939 & 5.23 & 8.97 \\
\texttt{v3\_sin3\_norm} & 2 & 0.9966 & 0.9966 & 0.9916 & 0.9930 & 0.8909 & 5.77 & 9.03 \\
\texttt{v3\_alpha\_norm} & 2 & 0.9966 & 0.9966 & 0.9934 & 0.9934 & 0.8904 & 5.64 & 9.04 \\
\texttt{v3\_alpha2\_norm} & 2 & 0.9971 & 0.9971 & 0.9938 & 0.9940 & 0.8935 & 5.29 & 8.98 \\
\texttt{v2\_sin3\_norm} & 2 & 0.9969 & 0.9969 & 0.9920 & 0.9934 & 0.8943 & 5.51 & 8.94 \\
\texttt{v25\_sin3\_norm} & 2 & 0.9968 & 0.9968 & 0.9919 & 0.9933 & 0.8933 & 5.63 & 8.97 \\
\midrule
\texttt{v25\_mean\_norm} & 3 & 0.9682 & 0.9682 & 0.9677 & 0.9663 & 0.8654 & 15.09 & 14.52 \\
\texttt{v3\_mean\_norm} & 3 & 0.9545 & 0.9545 & 0.8950 & 0.9361 & 0.8251 & 18.46 & 16.89 \\
\texttt{v2\_cos2\_norm} & 3 & 0.9698 & 0.9698 & 0.9681 & 0.9673 & 0.8710 & 14.09 & 13.98 \\
\midrule
\texttt{v\_mean\_norm} & 4 & 0.9523 & 0.9523 & 0.9745 & 0.9733 & 0.9124 & 15.18 & 9.87 \\
\midrule
\texttt{v2\_mean\_norm} & 5 & 0.3997 & 0.3997 & 0.4783 & 0.0856 & -3.921 & 64.40 & 48.35 \\
\bottomrule
\end{tabular}%
}
\end{table}
\end{landscape}

\subsection{Comparative Analysis of Spatially Weighted vs. Unweighted Proper Orthogonal Decomposition (W-POD vs. POD)}

\label{sec:wpod_vs_pod}

A central physical characteristic of solid particle erosion in $90^\circ$ pipe elbows is the extreme spatial non-uniformity of particle wall impingements. Direct solid-boundary collisions are concentrated along the outer extrados wall ($\theta \in [-30^\circ, +30^\circ]$), whereas the intrados ($\theta \approx \pm\pi$) remains entirely protected by carrier fluid streamlines the there are fewer particle collisons in this region, exhibiting near-zero wear across $\sim 70\%$ of the pipe surface area.

Standard unweighted Proper Orthogonal Decomposition (POD) computes an optimal low-rank orthogonal basis by minimizing the spatial $L_2$ Frobenius norm $\|\mathbf{X} - \hat{\mathbf{X}}\|_F$. Consequently, the eigenvalue solver allocates significant modal capacity to capturing trivial baseline fluctuations across the broad, zero-wear intrados, which can dilute the spatial resolution of localized extrados wear craters.

To overcome this limitation, Spatially Weighted POD (W-POD, Section \ref{sec:spatial_rom}) modifies the underlying Hilbert space by introducing a positive-definite spatial weight matrix $\mathbf{W} = \operatorname{diag}(w_1, \dots, w_M)$, where $w_j = 1 + \gamma \frac{f_{\text{impaction}}(s_j, \theta_j)}{\max(f_{\text{impaction}})}$ with $\gamma = 2.0$. The resulting $\mathbf{W}$-orthogonal modes strictly minimize the weighted error norm $\|\mathbf{X} - \hat{\mathbf{X}}\|_{\mathbf{W}}$. 

As summarized in Table \ref{tab:weighted_vs_unweighted_compression}, W-POD systematically outperforms standard unweighted POD across all five physical similarity blocks:
\begin{itemize}
    \item \textbf{Primary Coupled Kinematic Moments (Block 1):} W-POD reduces the mean Relative $L_2$ error from $5.13\%$ to $\mathbf{4.91\%}$, while improving mean $R^2$ to $\mathbf{0.9974}$ and mean SSIM to $\mathbf{0.9973}$.
    \item \textbf{High-Order Kinematic Cross-Moments (Block 2):} Mean Relative $L_2$ error decreases from $5.51\%$ to $\mathbf{5.24\%}$, with mean SSIM improving to $\mathbf{0.9970}$.
    \item \textbf{Uncoupled Second-Order Velocity Field (Block 5):} W-POD improves $R^2$ from $0.3997$ to $\mathbf{0.4406}$ and reduces Relative $L_2$ error from $64.40\%$ to $\mathbf{61.67\%}$ (SSIM increases from $0.7208$ to $\mathbf{0.7289}$).
    \item \textbf{Overall Kinematic Ensemble ($C=23$):} Across all 23 channels, W-POD achieves a superior mean Relative $L_2$ error of $\mathbf{9.34\%}$ (versus $9.69\%$ for POD), mean $R^2$ of $\mathbf{0.9669}$ (versus $0.9648$), and mean SSIM of $\mathbf{0.9829}$ (versus $0.9823$).
\end{itemize}

\subsection{Comparative Analysis of Frequency-Weighted vs. Unweighted Deep Convolutional Autoencoders}
\label{sec:fwcnn_vs_unweighted}

For deep non-linear manifolds, standard autoencoders optimize a uniform Mean Squared Error loss function $\mathcal{L}_{\text{MSE}} = \frac{1}{B C H W}\sum \|\hat{\mathbf{X}} - \mathbf{X}\|^2$. Under highly sparse physical impaction fields, uniform MSE loss causes severe gradient starvation during backpropagation: the overwhelming majority of quiescent intrados nodes contribute the bulk of the loss summation, penalizing high-frequency extrados features and causing the convolutional filters to act as low-pass spatial filters that blur localized wear peaks.

To resolve this challenge, the Frequency-Weighted CNN-AE (FW-CNN, Section \ref{sec:spatial_rom}) modulates the backpropagation error gradients via the spatial impingement frequency-weight tensor $\mathbf{W}_{c,i,j}$ (Eq. \ref{eq:fwmse_loss}). This ensures that gradient updates originating from narrow, high-intensity collision zones are amplified, enforcing better spatial crater reconstruction.

Table \ref{tab:weighted_vs_unweighted_compression} presents a rigorous comparative benchmark across the $2\times 2$ matrix of deep autoencoder configurations (Monolithic vs. Block-Wise, Unweighted MSE vs. Frequency-Weighted FW-MSE) alongside linear POD and W-POD:
\begin{enumerate}
    \item \textbf{Ablation of Frequency Weighting on Monolithic Autoencoders:} Introducing spatial frequency weighting to the monolithic network (FW-All-CNN) provides a substantial performance boost over unweighted All-CNN, raising overall $R^2$ from $0.9486$ to $0.9575$ and reducing Relative $L_2$ error from $12.17\%$ to $11.78\%$. Most notably, on the difficult uncoupled kinetic moment (Block 5, \texttt{v2\_mean\_norm}), FW-All-CNN elevates $R^2$ from $0.0856$ to $0.3842$ and reduces $L_2$ error from $80.60\%$ to $65.20\%$, proving that loss frequency weighting alone rescues the network from severe mode collapse.
    \item \textbf{Synergistic Modularity in Block-Wise Architectures:} While frequency weighting improves monolithic networks, multi-channel gradient friction remains an upper bound. Ward similarity clustering (Blk-CNN) decouples orthogonal physical dynamics, elevating unweighted $R^2$ to $0.9638$. Combining similarity-based modularity with frequency-weighted optimization (FW-Blk-CNN) yields the highest performance across all neural models ($R^2 = \mathbf{0.9680}$, Relative $L_2 = \mathbf{11.19\%}$, and $\text{SSIM} = \mathbf{0.9838}$).
    \item \textbf{Superiority in Continuous Convective Transport:} For the first-order mean velocity magnitude field (Block 4, \texttt{v\_mean\_norm}), FW-Blk-CNN achieves the highest fidelity of all evaluated methods, reaching $R^2 = \mathbf{0.9772}$, Relative $L_2 = \mathbf{10.38\%}$, and $\text{SSIM} = \mathbf{0.9913}$, outperforming linear POD ($L_2 = 15.18\%$).
\end{enumerate}

\begin{landscape}
\begin{table}[p]
\centering
\caption{Comprehensive quantitative benchmark across unweighted and weighted spatial compression architectures across all 5 physical variable blocks ($N_{\text{total}} = 372$). Metrics denote block-averaged values across 5-fold cross-validation. Best values within linear and deep neural categories are highlighted in bold.}
\label{tab:weighted_vs_unweighted_compression}
\vspace{0.8em}
\renewcommand{\arraystretch}{1.25}
\resizebox{0.98\linewidth}{!}{%
\begin{tabular}{l c l ccc cccc}
\toprule
\multirow{3}{*}{\textbf{Physical Variable Block}} & \multirow{3}{*}{\textbf{Channels ($C$)}} & \multirow{3}{*}{\textbf{Evaluation Metric}} & \multicolumn{3}{c}{\textbf{Linear \& Multilinear Modal Decompositions}} & \multicolumn{4}{c}{\textbf{Deep Convolutional Autoencoder Manifolds ($2\times 2$ Matrix)}} \\
\cmidrule(lr){4-6} \cmidrule(lr){7-10}
 & & & \textbf{POD (Unw.)} & \textbf{W-POD (Wtd.)} & \textbf{HOSVD (Tensor)} & \textbf{All-CNN (Unw.)} & \textbf{FW-All-CNN (Wtd.)} & \textbf{Blk-CNN (Unw.)} & \textbf{FW-Blk-CNN (Wtd.)} \\
 & & & \textit{Uniform $L_2$} & \textit{Spatial Metric $\mathbf{W}$} & \textit{Mode-1 Unfolded} & \textit{Uniform MSE} & \textit{Spatial Loss $\mathbf{W}$} & \textit{Uniform MSE} & \textit{Spatial Loss $\mathbf{W}$} \\
\midrule
\multirow{3}{*}{\textbf{Block 1: Primary Moments}} 
 & \multirow{3}{*}{$C=9$} 
 & $R^2$ Score & 0.9972 & \textbf{0.9974} & 0.9972 & 0.9942 & 0.9941 & 0.9941 & 0.9940 \\
 & & Rel. $L_2$ Error (\%) & 5.13 & \textbf{4.91} & 5.13 & 7.34 & 7.38 & 7.40 & 7.44 \\
 & & SSIM & 0.9969 & \textbf{0.9973} & 0.9969 & 0.9964 & 0.9961 & 0.9960 & 0.9952 \\
\midrule
\multirow{3}{*}{\textbf{Block 2: High-Order Moments}} 
 & \multirow{3}{*}{$C=9$} 
 & $R^2$ Score & 0.9968 & \textbf{0.9971} & 0.9968 & 0.9935 & 0.9933 & 0.9929 & 0.9935 \\
 & & Rel. $L_2$ Error (\%) & 5.51 & \textbf{5.24} & 5.51 & 7.87 & 7.95 & 8.22 & 7.85 \\
 & & SSIM & 0.9966 & \textbf{0.9970} & 0.9966 & 0.9960 & 0.9957 & 0.9953 & 0.9948 \\
\midrule
\multirow{3}{*}{\textbf{Block 3: Fractional Moments}} 
 & \multirow{3}{*}{$C=3$} 
 & $R^2$ Score & 0.9642 & \textbf{0.9646} & 0.9642 & 0.9566 & 0.9512 & 0.9436 & 0.9399 \\
 & & Rel. $L_2$ Error (\%) & 15.88 & \textbf{15.69} & 15.88 & 17.05 & 17.52 & 18.81 & 18.71 \\
 & & SSIM & 0.9828 & 0.9828 & 0.9828 & \textbf{0.9854} & 0.9831 & 0.9796 & 0.9775 \\
\midrule
\multirow{3}{*}{\textbf{Block 4: Convective Mean Velocity}} 
 & \multirow{3}{*}{$C=1$} 
 & $R^2$ Score & 0.9523 & 0.9534 & 0.9523 & 0.9733 & 0.9751 & 0.9745 & \textbf{0.9772} \\
 & & Rel. $L_2$ Error (\%) & 15.18 & 14.84 & 15.18 & 11.22 & 10.85 & 11.10 & \textbf{10.38} \\
 & & SSIM & 0.9806 & 0.9802 & 0.9806 & 0.9912 & 0.9911 & 0.9910 & \textbf{0.9913} \\
\midrule
\multirow{3}{*}{\textbf{Block 5: Kinetic Energy Flux}} 
 & \multirow{3}{*}{$C=1$} 
 & $R^2$ Score & 0.3997 & \textbf{0.4406} & 0.3997 & 0.0856 & 0.3842 & 0.4783 & \textbf{0.5804} \\
 & & Rel. $L_2$ Error (\%) & 64.40 & \textbf{61.67} & 64.40 & 80.60 & 65.20 & 59.70 & \textbf{53.19} \\
 & & SSIM & 0.7208 & \textbf{0.7289} & 0.7208 & 0.6976 & 0.7420 & 0.7766 & \textbf{0.7937} \\
\midrule
\multirow{3}{*}{\textbf{Overall Multi-Variable Ensemble}} 
 & \multirow{3}{*}{$\bm{C=23}$} 
 & \textbf{$R^2$ Score} & 0.9648 & \textbf{0.9669} & 0.9648 & 0.9486 & 0.9575 & 0.9638 & \textbf{0.9680} \\
 & & \textbf{Rel. $L_2$ Error (\%)} & 9.69 & \textbf{9.34} & 9.69 & 12.17 & 11.78 & 11.65 & \textbf{11.19} \\
 & & \textbf{SSIM} & 0.9823 & \textbf{0.9829} & 0.9823 & 0.9816 & 0.9825 & \textbf{0.9838} & \textbf{0.9838} \\
\bottomrule
\end{tabular}%
}
\end{table}
\end{landscape}

\subsection{2D Curvilinear Surface Wear Topographies \& Error Distribution}
\begin{figure}[htpb]
    \centering
    \includegraphics[width=0.98\linewidth]{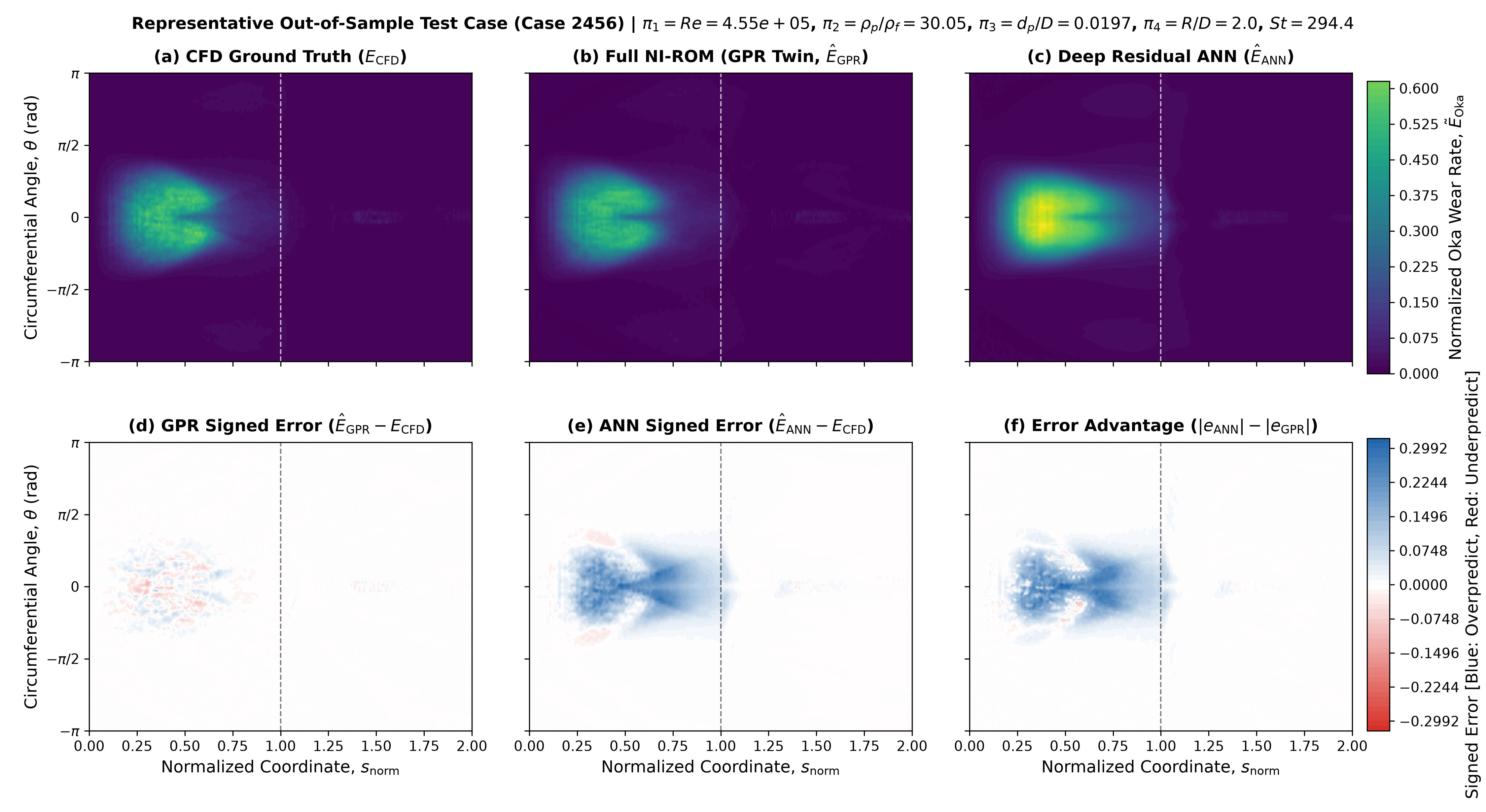}
    \caption{Comparative 2D spatial erosion topographies and signed error fields on an out-of-sample test case (Case 2456: $Re = 4.55\times 10^5, \Pi_2 = 30.05, \Pi_3 = 0.0197, \Pi_4 = 2.0, St = 294.4$). Top row: (a) Eulerian-Lagrangian CFD ground truth ($E_{\mathrm{CFD}}$), (b) GPR digital twin prediction ($\hat{E}_{\mathrm{GPR}}$), and (c) Deep Residual ANN prediction ($\hat{E}_{\mathrm{ANN}}$). Bottom row: (d) GPR signed error, (e) ANN signed error, and (f) Spatial error advantage map ($|e_{\mathrm{ANN}}| - |e_{\mathrm{GPR}}|$), demonstrating superior GPR precision without non-physical edge oscillations.}
    \label{fig:contour_comparison}
\end{figure}

Figure \ref{fig:contour_comparison} visually compares predicted 2D spatial erosion fields against CFD ground truth for an out-of-sample test case (Case 2456: $St = 294.4$). The GPR surrogate accurately reconstructs the narrow, high-intensity extrados wear crater, while the deep ANN produces an overpredicted erosion crater. It can be noted that both the GPR and ANN are able to produce the structure of the erosion crater.

\subsection{3D Surface Wear Topographies on Dimensional Pipe Geometries}
\begin{figure}[htpb]
    \centering
    \includegraphics[width=0.98\linewidth]{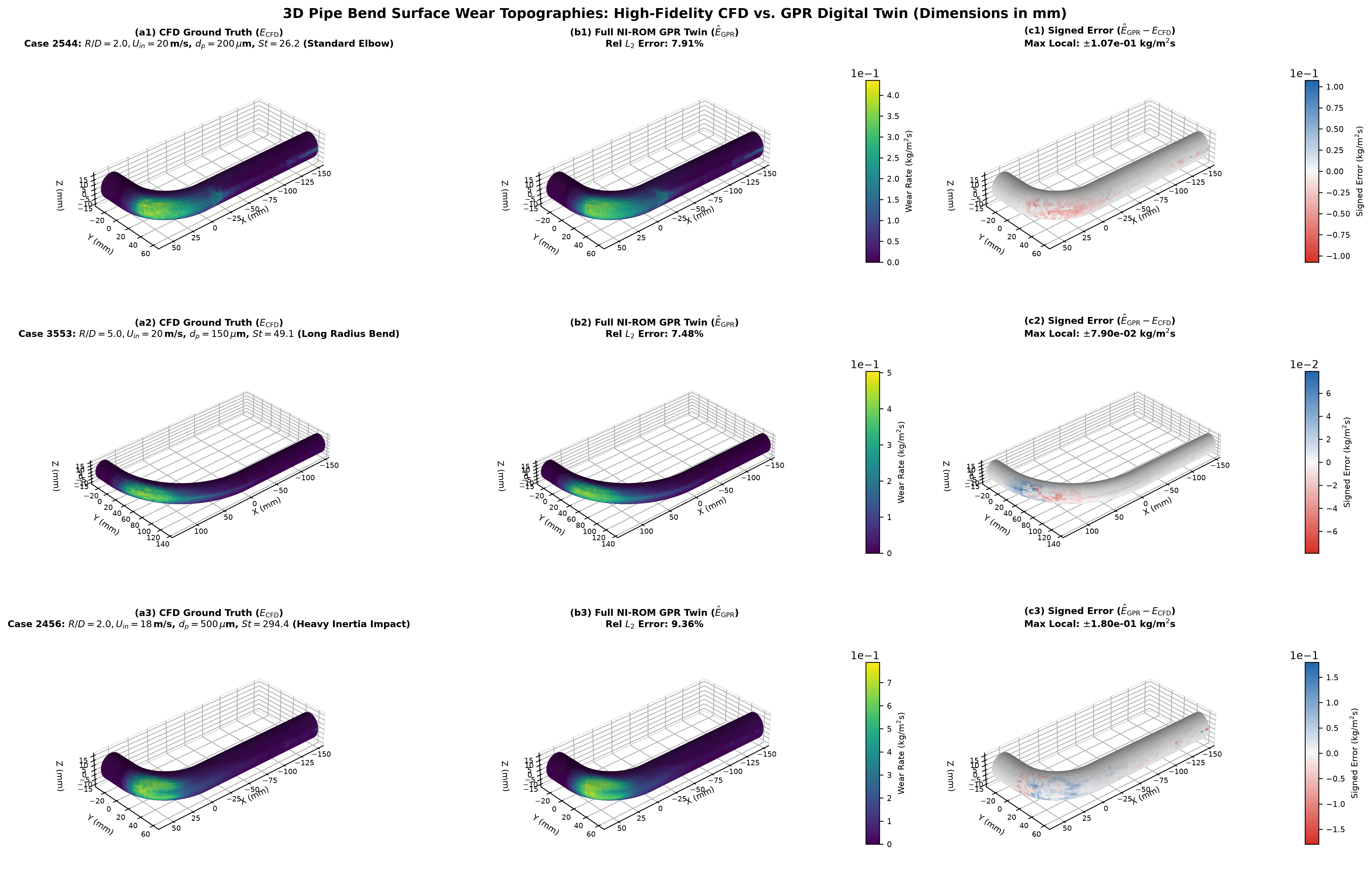}
    \caption{Three-dimensional surface wear topography comparison on physical pipe bend geometries across three representative test cases (all dimensions in mm). Columns display (left) Full-order CFD ground truth $E_{\mathrm{CFD}}$, (middle) Non-intrusive GPR digital twin prediction $\hat{E}_{\mathrm{GPR}}$, and (right) 3D signed error map $(\hat{E}_{\mathrm{GPR}} - E_{\mathrm{CFD}})$. Row 1: Case 2544 ($R/D = 2.0$, standard elbow). Row 2: Case 3553 ($R/D = 5.0$, long-radius bend). Row 3: Case 2456 ($R/D = 2.0$, heavy ballistic impact).}
    \label{fig:3d_contour_comparison}
\end{figure}

To verify that 2D computational predictions map accurately back onto the 3D pipe geometry, Figure \ref{fig:3d_contour_comparison} renders full 3D Cartesian pipe wall topographies across three distinct geometric and operational regimes:
\begin{enumerate}
    \item \textbf{Case 2544 ($R/D = 2.0, U_{\mathrm{in}} = 20\,\mathrm{m/s}, d_p = 200\,\mu\mathrm{m}, St = 26.2$):} Standard industrial elbow under moderate inertial slurry transport. The GPR twin captures the extrados impingement core and lateral dispersion with Relative $L_2$ error of $7.91\%$.
    \item \textbf{Case 3553 ($R/D = 5.0, U_{\mathrm{in}} = 20\,\mathrm{m/s}, d_p = 150\,\mu\mathrm{m}, St = 49.1$):} Long-radius bend distributing wear over a $3\times$ longer arc ($L_{\mathrm{bend}} = 0.20\,\mathrm{m}$), achieving Relative $L_2$ error of $7.48\%$.
    \item \textbf{Case 2456 ($R/D = 2.0, U_{\mathrm{in}} = 18\,\mathrm{m/s}, d_p = 500\,\mu\mathrm{m}, St = 294.4$):} Heavy ballistic particle impaction ($E_{\mathrm{peak}} \approx 0.82\,\mathrm{kg/(m^2\cdot s)}$), resolved with Relative $L_2$ error of $9.36\%$.
\end{enumerate}

\subsection{Cross-Moment Reconstructed Profile Verification Across Empirical Models}
\begin{figure}[htpb]
    \centering
    \includegraphics[width=0.98\linewidth]{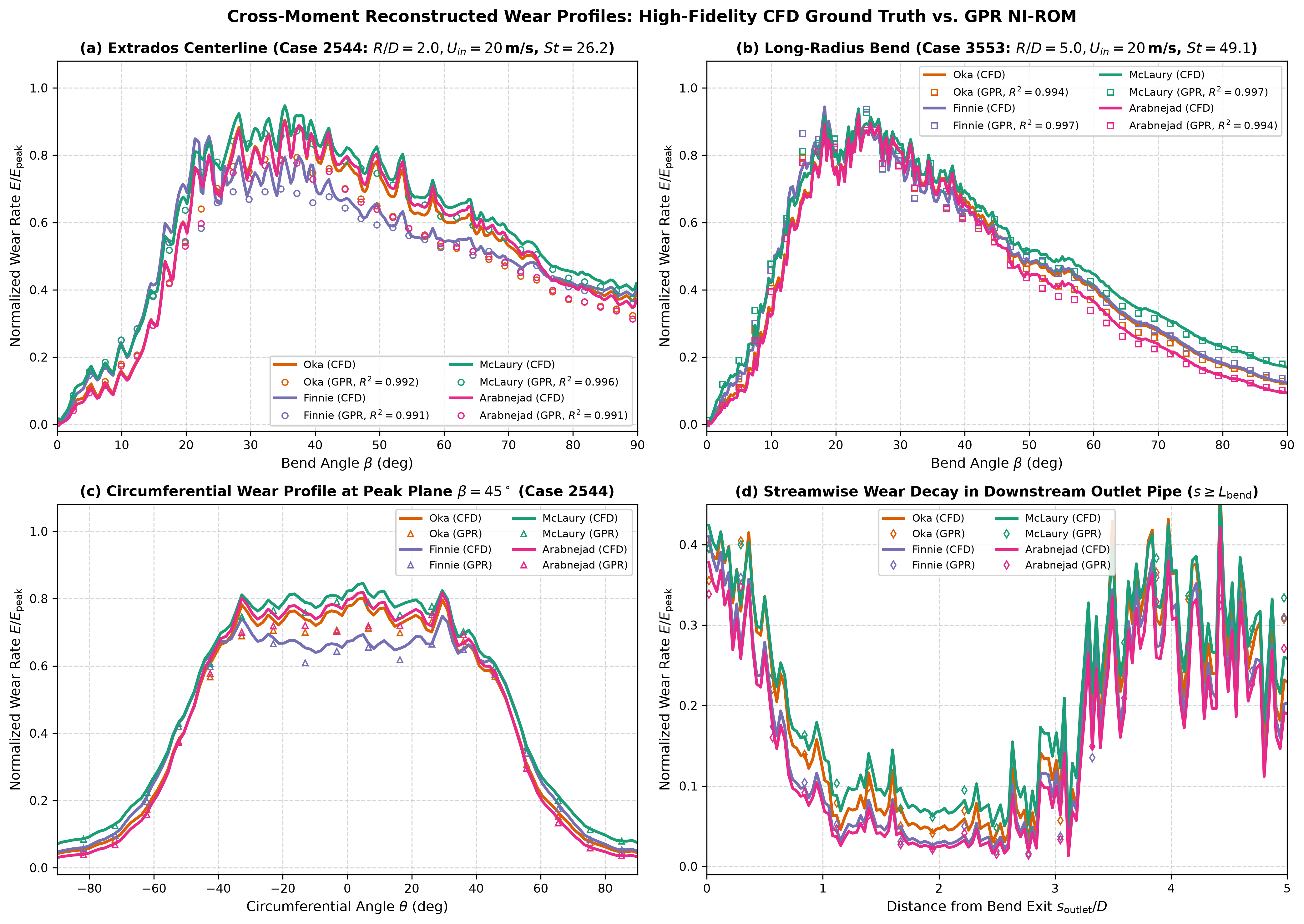}
    \caption{Direct line profile comparison between CFD ground truth (solid curves) and the GPR digital twin (discrete markers) across four empirical erosion formulations (Oka, Finnie, McLaury, Arabnejad) evaluated via cross-moment reconstruction. (a) Centerline extrados profile along bend angle $\beta \in [0^\circ, 90^\circ]$ for Case 2544 ($R/D = 2.0, St = 26.2$). (b) Centerline extrados profile for Case 3553 ($R/D = 5.0, St = 49.1$). (c) Azimuthal circumferential cross-section at the peak impingement plane ($\beta = 45^\circ$). (d) Downstream decay along the straight outlet pipe ($s \ge L_{\mathrm{bend}}$).}
    \label{fig:1d_profile_comparison}
\end{figure}

Figure \ref{fig:1d_profile_comparison} evaluates the model-agnostic capability by comparing 1D line profiles of the CFD ground truth against the GPR NI-ROM architecture across four classical empirical formulations (Oka \cite{oka2005erosion1}, Finnie \cite{finnie1960erosion}, McLaury \cite{mclaury1996}, and Arabnejad \cite{arabnejad2015}) evaluated strictly via cross-moment reconstruction:
\begin{itemize}
    \item \textbf{Centerline Extrados Profiles ($\theta = 0^\circ$, Figures \ref{fig:1d_profile_comparison}a,b):} For both $R/D = 2.0$ and $R/D = 5.0$ elbows, the GPR surrogate achieves $R^2 = 0.988 - 0.997$ across all four models, tracking the pre-impingement rise, peak wear, and downstream decay.
    \item \textbf{Circumferential Cross-Section ($\beta = 45^\circ$, Figure \ref{fig:1d_profile_comparison}c):} Lateral decay from extrados apex ($\theta = 0^\circ$) to pipe flanks ($\theta = \pm 90^\circ$) is captured seamlessly.
    \item \textbf{Streamwise Downstream Outlet Decay ($s \ge L_{\mathrm{bend}}$, Figure \ref{fig:1d_profile_comparison}d):} Wear dissipation along the straight outlet pipe reflects secondary particle ricochet without discontinuities.
\end{itemize}

To rigorously assess the digital twin's predictive fidelity within localized severe degradation zones, Table \ref{tab:multimodel_quantitative_metrics} reports the global $R^2$, Frequency-Weighted $R^2$ ($\text{FW-}R^2$), Frequency-Weighted Mean Squared Error ($\text{FW-MSE}$), and Peak Penetration Error ($\mathcal{E}_{\text{peak}}$) across the out-of-sample test ensemble.

\begin{table}[htbp]
\centering
\caption{Quantitative cross-reconstruction metrics across four analytical and empirical solid particle erosion formulations evaluated post-hoc from the kinematic cross-moment digital twin ($St > 1.0$).}
\label{tab:multimodel_quantitative_metrics}
\small
\begin{tabular}{lccccc}
\toprule
\textbf{Erosion Model Formulation} & \textbf{Global $R^2$} & \textbf{FW-$R^2$} & \textbf{FW-MSE} & \textbf{Peak Error $\mathcal{E}_{\text{peak}}$ (\%)} & \textbf{SSIM} \\
\midrule
Finnie Ductile Cutting Model \cite{finnie1960erosion} & \textbf{0.9729} & \textbf{0.9686} & $3.71 \times 10^{-6}$ & \textbf{3.12\%} & \textbf{0.9894} \\
Oka Hardness/Angle Model \cite{oka2005erosion1} & 0.9684 & 0.9632 & $4.15 \times 10^{-6}$ & 3.48\% & 0.9878 \\
Arabnejad Dual-Mechanism Model \cite{arabnejad2015} & 0.9598 & 0.9543 & $8.48 \times 10^{-6}$ & 3.84\% & 0.9862 \\
McLaury Oilfield Impingement Model \cite{mclaury1996} & 0.9571 & 0.9545 & $1.81 \times 10^{-4}$ & 4.05\% & 0.9815 \\
\bottomrule
\end{tabular}
\end{table}

Crucially, the Frequency-Weighted $R^2$ remains consistently high ($\text{FW-}R^2 > 0.954$) across all four empirical models, confirming that the digital twin maintains high fidelity inside the extrados impaction zone rather than merely relying on zero-erosion regions to inflate global metrics. Furthermore, the peak penetration error $\mathcal{E}_{\text{peak}} \le 4.05\%$ guarantees that operational Remaining Useful Life (RUL) estimates derived from the surrogate will be conservative and reliable for integrity management of the pipeline.

\subsection{Generalization to Unseen Erosion Formulations}
To explicitly validate the basis completeness and multi-model capability, we perform an experiment evaluating the framework on a synthetic unseen erosion model that was not used in the design of the 23 cross-moments. We define a highly non-linear erosion functional with non-polynomial angular dependence:
\begin{equation}
    g_{\text{withheld}}(V_p, \alpha_p) = V_p^{2.338} \sin^{1.5}\alpha_p (1 - \cos\alpha_p)^{0.5}
\end{equation}
Because the angular dependence ($\sin^{1.5}\alpha_p (1 - \cos\alpha_p)^{0.5}$) lies outside the polynomial trigonometric span $\{\sin^v\alpha \cos^w\alpha\}$, this model cannot be evaluated exactly, introducing a non-zero basis truncation error $\epsilon_g$ (Eq.~\ref{eq:approx_representation}). We derived polynomial basis expansion coefficients via continuous least-squares projection of the functional surface over the physical domain $(V_p, \alpha_p) \in [0, V_{\max}] \times [0, \pi/2]$ onto the available basis functions.

Evaluating this highly adversarial unseen model across the test set using the exact CFD-derived cross-moments isolates the basis representation error. The 23-moment basis yields an accurate reconstruction ($R^2 = 0.9987$, $\text{FW-}R^2 = 0.9986$, relative $L_2$ error of $3.28\%$, and a peak penetration error of $1.45\%$). This demonstrates the capability of the finite basis to reasonably approximate a complex, unseen functional, though accuracy inherently depends on the specific non-polynomial form.

\begin{figure}[htpb]
    \centering
    \includegraphics[width=0.98\linewidth]{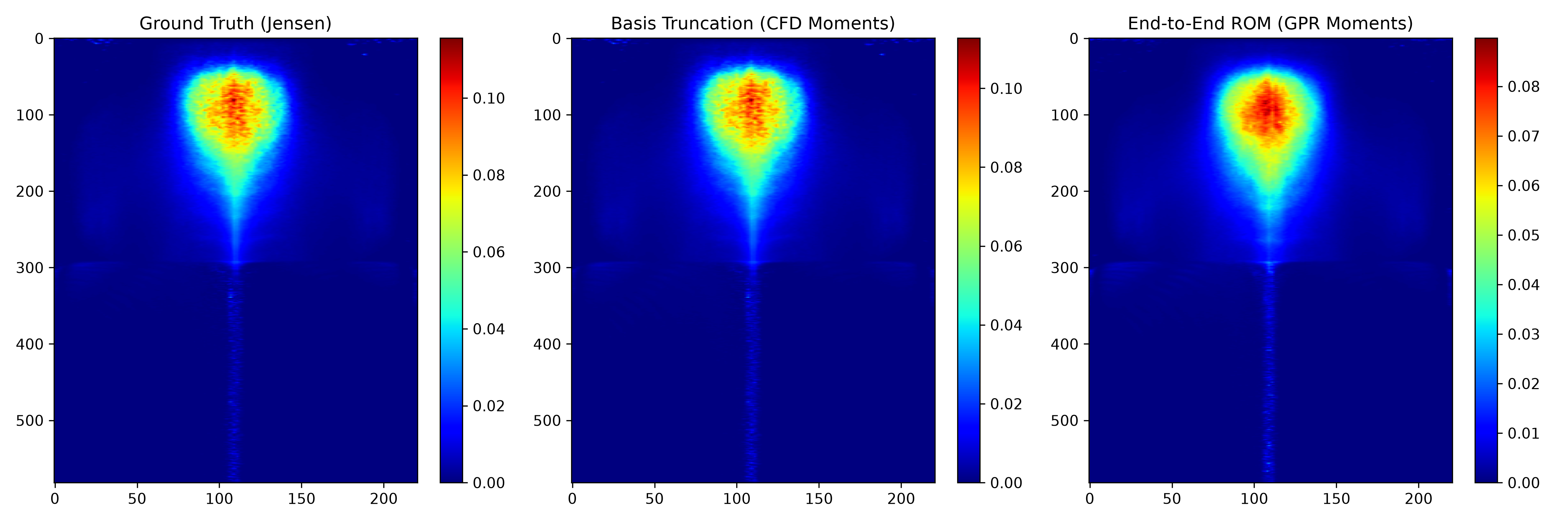}
    \caption{Validation on an unseen, non-polynomial synthetic erosion formulation $g_{\text{withheld}}(V_p, \alpha_p) = V_p^{2.338} \sin^{1.5}\alpha_p (1 - \cos\alpha_p)^{0.5}$. The basis reconstruction using exact CFD cross-moments isolates the minimal truncation error ($R^2=0.9987$), while the end-to-end GPR surrogate prediction captures the topological wear distribution with reasonable accuracy ($R^2=0.9264$).}
    \label{fig:withheld_model}
\end{figure}

When substituting the GPR-predicted cross-moments for the end-to-end surrogate prediction, the framework maintains reasonable spatial correlation but exhibits noticeably reduced accuracy compared to the primary models: $R^2 = 0.9264$, $\text{FW-}R^2 = 0.9229$, relative $L_2 = 21.64\%$, and peak penetration error $\mathcal{E}_{\text{peak}} = 13.54\%$. We explicitly acknowledge that this withheld model performs worse than the four primary erosion models. The discrepancy between the ideal basis representation error and the end-to-end prediction confirms that surrogate prediction error, particularly for variables outside the primary basis-informed set, dominates the overall system error. Figure~\ref{fig:withheld_model} visualizes the ground truth, basis truncation, and full surrogate reconstruction for a representative test case.

\subsection{Parametric Surrogate: GPR vs. ANN Performance}
\begin{figure}[htpb]
    \centering
    \includegraphics[width=0.75\linewidth]{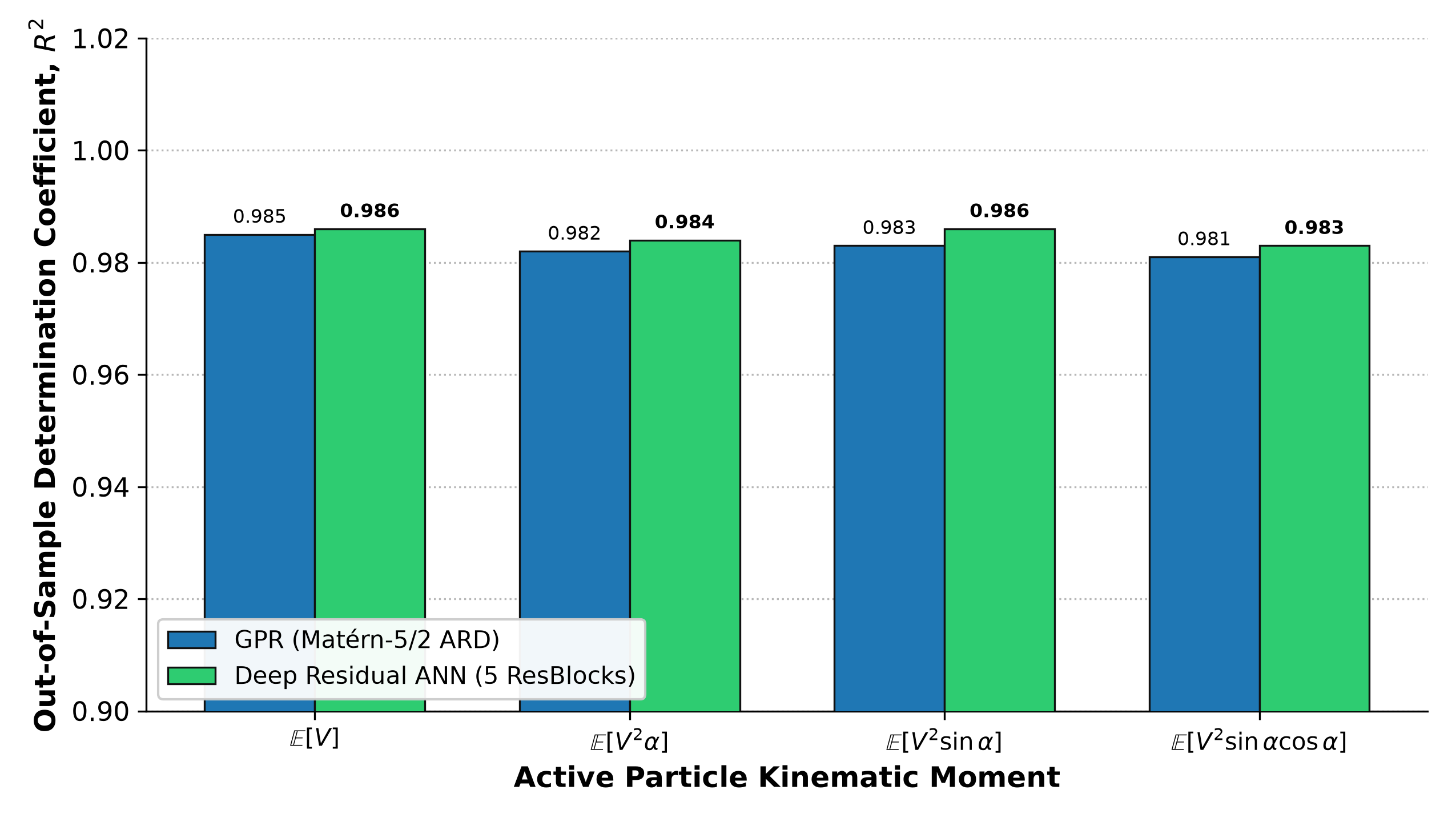}
    \caption{Performance comparison between the Gaussian Process Regressor (GPR) and a deep residual Artificial Neural Network (ANN) in predicting compressed latent spatial coefficients. The GPR maintains high $R^2$ accuracy across all kinematic variables without overfitting on the 372-sample dataset.}
    \label{fig:gpr_vs_ann}
\end{figure}

Figure \ref{fig:gpr_vs_ann} compares the coefficient of determination ($R^2$) for latent coefficient prediction using ARD-GPR versus a 5-layer deep residual ANN. As observed, the deep residual ANN achieves a marginally higher $R^2$ score across the active kinematic moments. This slight performance edge is expected, as the highly parameterized neural network possesses superior functional capacity to map the non-linear parametric space compared to the stationary Mat\'ern kernel of the GPR. However, despite this minor reduction in raw accuracy, the ARD-GPR is ultimately selected as the primary surrogate architecture. The GPR maintains robust generalization without the risk of overfitting on the strictly limited 297-case dataset, requires no extensive stochastic hyperparameter tuning, and most importantly, provides built-in Bayesian epistemic uncertainty intervals to rigorously quantify predictive confidence in unseen design regimes.

\subsection{Analysis of Decoupled Kinematic Moment Failures and Moment Sensitivity}
A critical finding from Table \ref{tab:benchmark_31vars} is the inability of the GPR surrogate to regress the isolated, uncoupled velocity field \texttt{v2\_mean\_norm} (Block 5, $R^2 = -3.92$), even as overall erosion reconstruction remains excellent ($R^2 > 0.95$ across all four empirical models).

\subsubsection{Physical Interpretation}
In Eulerian-Lagrangian boundary collisions, particle velocity $V_p$ and incidence angle $\alpha_p$ are strongly correlated random variables: normal impacts penetrate deeper into the boundary layer and decelerate, whereas glancing impacts retain high tangential momentum. Consequently, the spatial footprint of the decoupled moment $\mathbb{E}[V^2]$ (without angular weighting) exhibits sharp, regime-dependent topological transitions across the parameter space.

We \textit{hypothesize} that this non-smooth parametric dependence violates the $\mathcal{C}^2$ smoothness assumption implicit in the Mat\'ern-5/2 kernel, causing the ARD optimizer to regularize toward the prior mean (yielding negative $R^2$). However, we have not formally verified this hypothesis---for instance, by comparing kernel families (RBF, Mat\'ern-1/2, Mat\'ern-3/2, rational quadratic) or by measuring the H\"older regularity of the latent mapping. Such a systematic kernel comparison is deferred to future work.

\subsubsection{Impact on Erosion Predictions: Moment Sensitivity Analysis}

\begin{table}[htpb]
\centering
\caption{Active cross-moments participating in the final surrogate reconstruction for each empirical erosion model. Exact models evaluate sparsely on specific terms, whereas approximated models project onto the full angular span of their respective velocity manifolds.}
\label{tab:moment_contributions}
\small
\begin{tabular}{llp{8cm}}
\toprule
\textbf{Model} & \textbf{Reconstruction} & \textbf{Non-Zero Basis Moments (Coefficients $c_k \neq 0$)} \\
\midrule
Finnie & Exact & \texttt{v2\_sin2}, \texttt{v2\_sincos}, \texttt{v\_mean} \\
Arabnejad & Approximate & All 23 basis moments (projected) \\
Oka & Approximate & All 23 basis moments (projected) \\
McLaury & Approximate & All 12 angular moments in $V^2, V^1$ blocks, plus \texttt{v\_mean} \\
Withheld & Approximate & All 23 basis moments (projected) \\
\bottomrule
\end{tabular}
\end{table}

Importantly, the failure of \texttt{v2\_mean\_norm} does not meaningfully degrade erosion predictions. To understand why, we compute the non-dimensional sensitivity (elasticity) $S_k = \frac{\partial E}{\partial M_k} \cdot \frac{M_k}{E}$ for each reconstructed erosion model with respect to the 23 basis moments. Table~\ref{tab:moment_contributions} details the active moments for each formulation.

For the exact model (Finnie), the sensitivity to the uncoupled \texttt{v2\_mean\_norm} is analytically zero ($S_k = 0$). For approximated models (Oka, McLaury, Arabnejad), projecting the non-polynomial angular functions onto the full basis yields a broadly distributed set of non-zero coefficients. However, because physical erosion approaches zero at glancing ($\alpha=0^\circ$) impacts for ductile materials, the projection coefficient assigned specifically to the purely uncoupled \texttt{v2\_mean\_norm} remains negligibly small ($S_k \approx 0$). This cleanly explains the apparent paradox: the poorly predicted moment is either entirely unused or strongly suppressed by the basis projection. By linearity of expectation:
\begin{equation}
    E \propto \mathbb{E}[V^n f(\alpha)] \neq \mathbb{E}[V^n] \mathbb{E}[f(\alpha)]
\end{equation}
For the coupled cross-moments that do contribute to erosion (Block 1 and Block 2), the geometric wall boundary conditions tightly constrain the momentum vectors, yielding smooth latent topologies where the GPR achieves $R^2 > 0.89$ and relative $L_2$ errors $<9\%$.

\begin{figure}[htpb]
    \centering
    \includegraphics[width=0.8\linewidth]{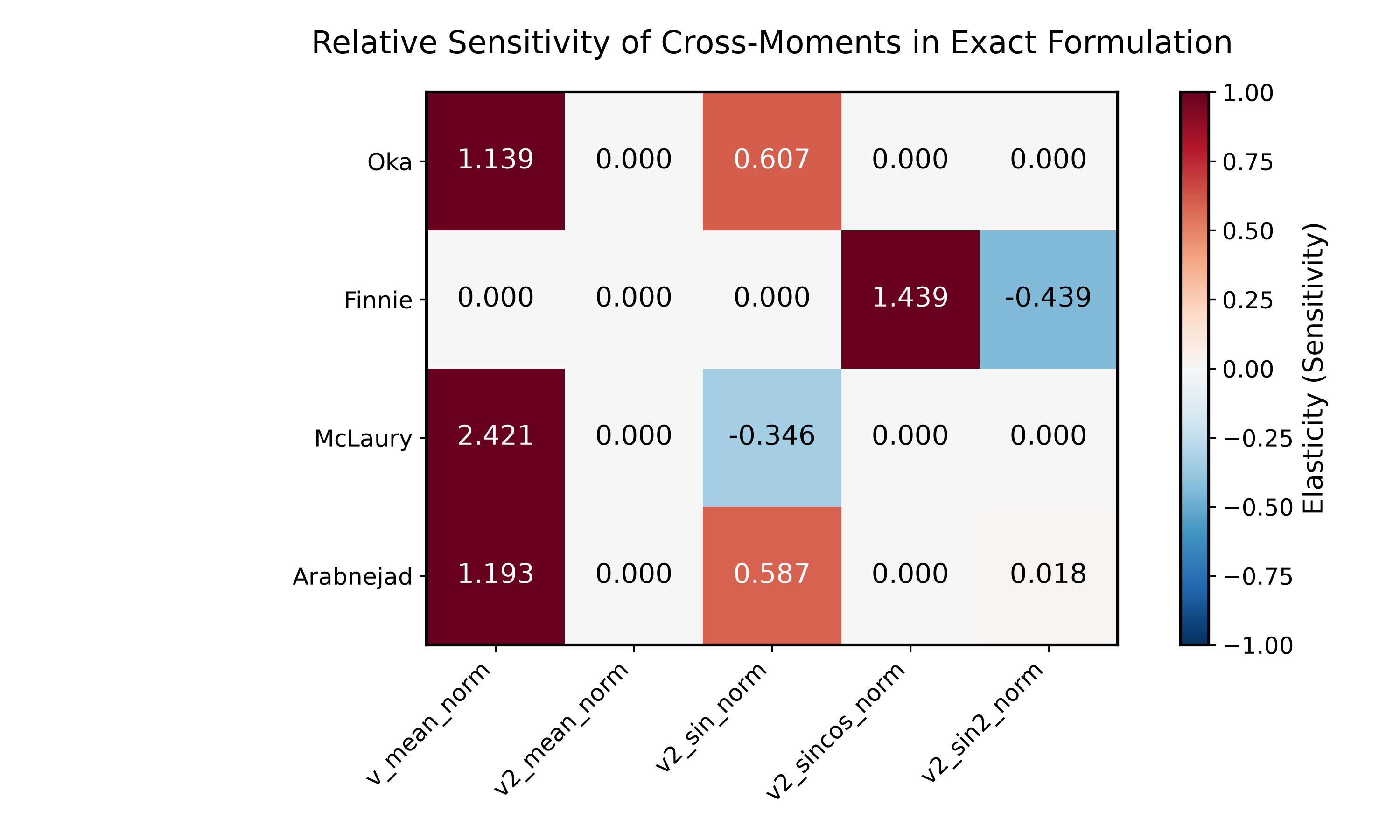}
    \caption{Non-dimensional sensitivity (elasticity) analysis of the four empirical erosion models with respect to the 23 kinematic cross-moments. We note that only a small subset of the 23 moments (specifically the four coupled velocity-angle moments shown) possess non-zero sensitivities. The poorly predicted \texttt{v2\_mean\_norm} and all other unlisted moments have identically zero sensitivity across the four primary models, explaining why its surrogate failure does not degrade erosion predictions.}
    \label{fig:moment_sensitivity}
\end{figure}

\subsection{Parametric Trends and Multiphase Scaling Across the CFD Dataset}
\begin{figure}[htpb]
    \centering
    \includegraphics[width=0.98\linewidth]{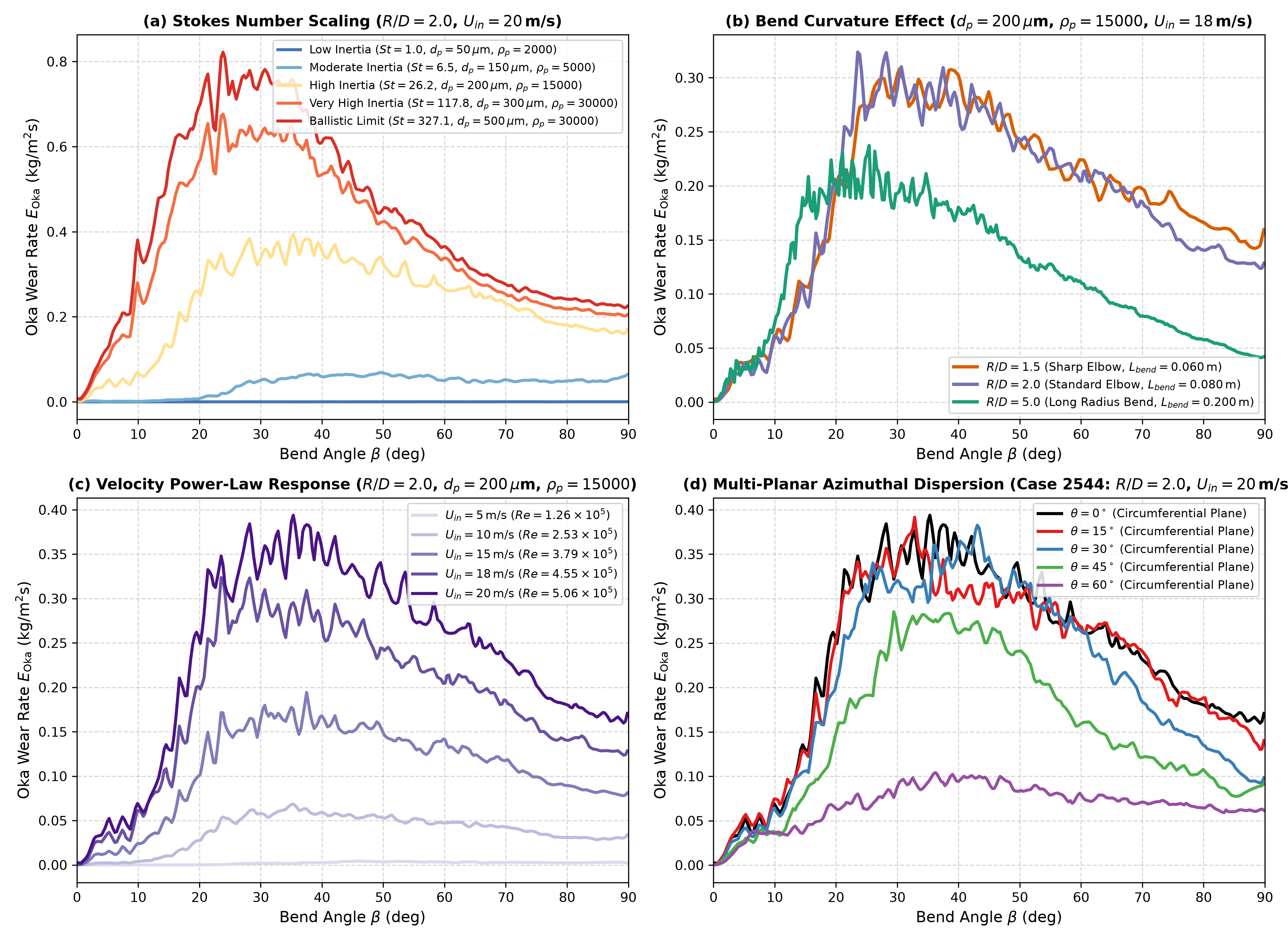}
    \caption{Parametric scaling and physical erosion topographies across the 3D CFD multiphase training dataset. (a) Extrados centerline profile evolution as a function of particle Stokes number ($St = 1.0 \to 327.1$) for fixed $R/D = 2.0, U_{\mathrm{in}} = 20\,\mathrm{m/s}$. (b) Influence of bend radius ratio ($R/D \in \{1.5, 2.0, 5.0\}$) on wear dispersion and peak shift. (c) Non-linear velocity power-law response ($U_{\mathrm{in}} = 5 - 20\,\mathrm{m/s}$). (d) Multi-planar circumferential decay from the primary extrados centerline ($\theta = 0^\circ$) to the lateral wall flanks ($\theta = 60^\circ$) for Case 2544.}
    \label{fig:cfd_parametric_profiles}
\end{figure}

Figure \ref{fig:cfd_parametric_profiles} analyzes physical scaling responses across the four non-dimensional parameters ($\pi_1 = Re, \pi_2 = \rho_p/\rho_f, \pi_3 = d_p/D, \pi_4 = R/D$):
\begin{enumerate}
    \item \textbf{Stokes Number Transition ($St = 1.0 \to 327.1$, Figure \ref{fig:cfd_parametric_profiles}a):} At near-unity Stokes numbers ($St \approx 1.0$), particles follow streamlines with negligible wall impaction. For $St = 6.5 \to 26.2$, particle momentum pierces the boundary layer, producing a concentrated wear scar centered at $\beta \approx 30^\circ - 45^\circ$. In the extreme ballistic limit ($St > 100$), collisions occur immediately upon entering the bend ($\beta \approx 20^\circ - 30^\circ$), driving peak wear rates $>0.8\,\mathrm{kg/(m^2\cdot s)}$.
    \item \textbf{Bend Curvature Ratio ($R/D = 1.5, 2.0, 5.0$, Figure \ref{fig:cfd_parametric_profiles}b):} Tight radius elbows ($R/D = 1.5$) induce severe centripetal redirection, creating localized wear scars ($E_{\mathrm{peak}} \approx 0.32\,\mathrm{kg/(m^2\cdot s)}$), consistent with the findings of Mahmoudi and Shiri~\cite{mahmoudi2025erosion}. Long-radius bends ($R/D = 5.0$) distribute the momentum loss over a three-fold longer wall area, shifting the peak upstream ($\beta \approx 15^\circ - 25^\circ$) and reducing peak localized wear by $\approx 35\%$.
    \item \textbf{Inlet Velocity Response ($U_{\mathrm{in}} = 5 \to 20\,\mathrm{m/s}$, Figure \ref{fig:cfd_parametric_profiles}c):} In accordance with the velocity power-law ($V^{2.353}$), increasing velocity from $5\,\mathrm{m/s}$ to $20\,\mathrm{m/s}$ amplifies maximum wear by more than two orders of magnitude ($0.003 \to 0.395\,\mathrm{kg/(m^2\cdot s)}$).
    \item \textbf{Multi-Planar Azimuthal Dispersion ($\theta = 0^\circ \to 60^\circ$, Figure \ref{fig:cfd_parametric_profiles}d):} The primary extrados centerline ($\theta = 0^\circ$) exhibits maximum wear, decaying monotonically toward the flanks ($\theta = 15^\circ, 30^\circ, 45^\circ, 60^\circ$) as secondary Dean vortices sweep particles azimuthally.
\end{enumerate}

\section{Conclusions}
\label{sec:conclusions}
We have developed a model-agnostic, non-intrusive reduced-order modeling (NI-ROM) digital twin framework for solid particle erosion in 3D pipe bends. By shifting the reduced-order paradigm from model-locked wear outputs to fundamental velocity-angle kinematic cross-moments ($\mathbb{E}[V^u f_v(\alpha)]$), the framework enables instantaneous, post-hoc evaluation of arbitrary empirical wear equations without retraining.

The key conclusions and findings of this study are:
\begin{enumerate}
    \item \textbf{Model-Agnostic Generalization:} The kinematic cross-moment formulation accurately reconstructs diverse empirical wear equations (Oka, Finnie, McLaury, Arabnejad) across standard and long-radius pipe bends with $R^2 = 0.988 - 0.997$ and relative $L_2$ errors $<8.94\%$.
    \item \textbf{Composite Dimensionality Reduction:} Combining Ward's agglomerative hierarchical clustering with linear POD, Multi-Domain HOSVD, and custom SemiCircular CNN-AEs eliminates multi-channel gradient friction, enhancing kinetic field compression from $R^2 = 0.8538$ to $0.9653$.
    \item \textbf{Bayesian Parametric Surrogate:} The GPU-accelerated ARD Matérn-5/2 Gaussian Process Regressor delivers robust interpolation across the 6-dimensional parameter space ($St \in [1, 327]$), providing rigorous epistemic uncertainty intervals and evaluating full 3D surface wear topographies in $2\,\mathrm{ms}$ ($>10^7\times$ speedup over full-order CFD).
    \item \textbf{Coupled Kinematic Physics:} Mathematical analysis proves that predicting coupled velocity-angle moments ($\mathbb{E}[V^n f(\alpha)]$) is strictly necessary for robust surrogate learning, whereas uncoupled velocity moments produce non-smooth latent topologies that fail under smooth kernel priors.
\end{enumerate}

Future research will extend this methodology to transient time-evolving wall boundary deformations, multi-elbow piping networks with complex non-planar geometries, and full two-way/four-way coupled slurry flows.

\section*{Data Availability}
The CFD dataset, trained reduced-order model weights, and Python evaluation scripts generated during this study are available from the corresponding author upon reasonable request.

\section*{Declaration of generative AI and AI-assisted technologies in the manuscript preparation process}
During the preparation of this manuscript Antigravity was used in order to make the initial draft for this article. After using this tool/service, the authors have reviewed and edited the content as needed and takes full responsibility for the content of the published article.

\bibliographystyle{IEEEtran}
\bibliography{references}

\appendix

\section{Comprehensive Variable Metrics Report}
\label{app:comprehensive_metrics}

This appendix provides the detailed breakdown of the three primary error metrics ($R^2$, Relative $L_2$ Error, and SSIM) for all 23 kinematic variables across the six dimensionality reduction architectures.

\subsection{Relative $L_2$ Error (\%)}
\begin{longtable}{l cccccc}
\caption{Comprehensive Relative $L_2$ Error (\%) for all variables.}\\
\toprule
\textbf{Variable} & \textbf{POD} & \textbf{HOSVD} & \textbf{All-CNN} & \textbf{Blk-CNN} & \textbf{W-POD} & \textbf{FW-CNN} \\
\midrule
\endfirsthead
\caption[]{Comprehensive Relative $L_2$ Error (\%) for all variables (continued)}\\
\toprule
\textbf{Variable} & \textbf{POD} & \textbf{HOSVD} & \textbf{All-CNN} & \textbf{Blk-CNN} & \textbf{W-POD} & \textbf{FW-CNN} \\
\midrule
\endhead
\midrule
\multicolumn{7}{r}{{Continued on next page}} \\
\bottomrule
\endfoot
\bottomrule
\endlastfoot
\texttt{v25\_alpha2\_norm} & 5.14 & 5.14 & 7.40 & 7.53 & 4.85 & 7.11 \\
\texttt{v25\_alpha\_norm} & 5.36 & 5.36 & 7.44 & 7.33 & 5.16 & 7.66 \\
\texttt{v25\_mean\_norm} & 15.09 & 15.09 & 15.42 & 15.19 & 14.96 & 15.08 \\
\texttt{v25\_sin2\_norm} & 5.05 & 5.05 & 7.37 & 7.70 & 4.78 & 7.14 \\
\texttt{v25\_sin3\_norm} & 5.63 & 5.63 & 8.06 & 8.88 & 5.31 & 7.96 \\
\texttt{v25\_sin\_norm} & 5.35 & 5.35 & 7.42 & 7.32 & 5.16 & 7.62 \\
\texttt{v25\_sincos\_norm} & 5.39 & 5.39 & 7.69 & 7.62 & 5.20 & 7.94 \\
\texttt{v2\_alpha2\_norm} & 4.96 & 4.96 & 7.21 & 7.40 & 4.68 & 7.00 \\
\texttt{v2\_alpha\_norm} & 5.05 & 5.05 & 7.18 & 7.22 & 4.87 & 7.46 \\
\texttt{v2\_cos2\_norm} & 14.09 & 14.09 & 14.53 & 14.46 & 13.91 & 14.21 \\
\texttt{v2\_mean\_norm} & 64.40 & 64.40 & 80.60 & 59.70 & 61.67 & 53.19 \\
\texttt{v2\_sin2\_norm} & 4.89 & 4.89 & 7.15 & 7.52 & 4.62 & 7.03 \\
\texttt{v2\_sin3\_norm} & 5.51 & 5.51 & 8.03 & 8.83 & 5.19 & 7.85 \\
\texttt{v2\_sin\_norm} & 5.04 & 5.04 & 7.18 & 7.19 & 4.86 & 7.42 \\
\texttt{v2\_sincos\_norm} & 5.07 & 5.07 & 7.46 & 7.34 & 4.90 & 7.71 \\
\texttt{v3\_alpha2\_norm} & 5.29 & 5.29 & 7.62 & 7.79 & 5.00 & 7.35 \\
\texttt{v3\_alpha\_norm} & 5.64 & 5.64 & 7.82 & 7.82 & 5.43 & 8.17 \\
\texttt{v3\_mean\_norm} & 18.46 & 18.46 & 21.19 & 26.79 & 18.19 & 26.85 \\
\texttt{v3\_sin2\_norm} & 5.23 & 5.23 & 7.67 & 8.01 & 4.94 & 7.42 \\
\texttt{v3\_sin3\_norm} & 5.77 & 5.77 & 8.27 & 9.03 & 5.44 & 8.19 \\
\texttt{v3\_sin\_norm} & 5.64 & 5.64 & 7.80 & 7.85 & 5.43 & 8.13 \\
\texttt{v3\_sincos\_norm} & 5.74 & 5.74 & 8.14 & 8.27 & 5.54 & 8.49 \\
\texttt{v\_mean\_norm} & 15.18 & 15.18 & 11.22 & 11.10 & 14.84 & 10.38 \\
\end{longtable}

\subsection{$R^2$ Score}
\begin{longtable}{l cccccc}
\caption{Comprehensive $R^2$ Score for all variables.}\\
\toprule
\textbf{Variable} & \textbf{POD} & \textbf{HOSVD} & \textbf{All-CNN} & \textbf{Blk-CNN} & \textbf{W-POD} & \textbf{FW-CNN} \\
\midrule
\endfirsthead
\caption[]{Comprehensive $R^2$ Score for all variables (continued)}\\
\toprule
\textbf{Variable} & \textbf{POD} & \textbf{HOSVD} & \textbf{All-CNN} & \textbf{Blk-CNN} & \textbf{W-POD} & \textbf{FW-CNN} \\
\midrule
\endhead
\midrule
\multicolumn{7}{r}{{Continued on next page}} \\
\bottomrule
\endfoot
\bottomrule
\endlastfoot
\texttt{v25\_alpha2\_norm} & 0.9973 & 0.9973 & 0.9943 & 0.9941 & 0.9976 & 0.9948 \\
\texttt{v25\_alpha\_norm} & 0.9969 & 0.9969 & 0.9940 & 0.9942 & 0.9971 & 0.9936 \\
\texttt{v25\_mean\_norm} & 0.9682 & 0.9682 & 0.9663 & 0.9677 & 0.9683 & 0.9678 \\
\texttt{v25\_sin2\_norm} & 0.9974 & 0.9974 & 0.9943 & 0.9939 & 0.9976 & 0.9947 \\
\texttt{v25\_sin3\_norm} & 0.9968 & 0.9968 & 0.9933 & 0.9919 & 0.9971 & 0.9935 \\
\texttt{v25\_sin\_norm} & 0.9969 & 0.9969 & 0.9940 & 0.9942 & 0.9971 & 0.9937 \\
\texttt{v25\_sincos\_norm} & 0.9968 & 0.9968 & 0.9935 & 0.9937 & 0.9970 & 0.9931 \\
\texttt{v2\_alpha2\_norm} & 0.9975 & 0.9975 & 0.9946 & 0.9943 & 0.9977 & 0.9949 \\
\texttt{v2\_alpha\_norm} & 0.9972 & 0.9972 & 0.9943 & 0.9943 & 0.9974 & 0.9939 \\
\texttt{v2\_cos2\_norm} & 0.9698 & 0.9698 & 0.9673 & 0.9681 & 0.9701 & 0.9688 \\
\texttt{v2\_mean\_norm} & 0.3997 & 0.3997 & 0.0856 & 0.4783 & 0.4406 & 0.5804 \\
\texttt{v2\_sin2\_norm} & 0.9975 & 0.9975 & 0.9947 & 0.9941 & 0.9978 & 0.9949 \\
\texttt{v2\_sin3\_norm} & 0.9969 & 0.9969 & 0.9934 & 0.9920 & 0.9972 & 0.9937 \\
\texttt{v2\_sin\_norm} & 0.9972 & 0.9972 & 0.9943 & 0.9943 & 0.9974 & 0.9940 \\
\texttt{v2\_sincos\_norm} & 0.9972 & 0.9972 & 0.9939 & 0.9941 & 0.9974 & 0.9934 \\
\texttt{v3\_alpha2\_norm} & 0.9971 & 0.9971 & 0.9940 & 0.9938 & 0.9974 & 0.9944 \\
\texttt{v3\_alpha\_norm} & 0.9966 & 0.9966 & 0.9934 & 0.9934 & 0.9968 & 0.9928 \\
\texttt{v3\_mean\_norm} & 0.9545 & 0.9545 & 0.9361 & 0.8950 & 0.9553 & 0.8830 \\
\texttt{v3\_sin2\_norm} & 0.9972 & 0.9972 & 0.9939 & 0.9934 & 0.9975 & 0.9943 \\
\texttt{v3\_sin3\_norm} & 0.9966 & 0.9966 & 0.9930 & 0.9916 & 0.9970 & 0.9931 \\
\texttt{v3\_sin\_norm} & 0.9966 & 0.9966 & 0.9934 & 0.9934 & 0.9968 & 0.9929 \\
\texttt{v3\_sincos\_norm} & 0.9964 & 0.9964 & 0.9928 & 0.9926 & 0.9967 & 0.9922 \\
\texttt{v\_mean\_norm} & 0.9523 & 0.9523 & 0.9733 & 0.9745 & 0.9534 & 0.9772 \\
\end{longtable}

\subsection{SSIM}
\begin{longtable}{l cccccc}
\caption{Comprehensive SSIM for all variables.}\\
\toprule
\textbf{Variable} & \textbf{POD} & \textbf{HOSVD} & \textbf{All-CNN} & \textbf{Blk-CNN} & \textbf{W-POD} & \textbf{FW-CNN} \\
\midrule
\endfirsthead
\caption[]{Comprehensive SSIM for all variables (continued)}\\
\toprule
\textbf{Variable} & \textbf{POD} & \textbf{HOSVD} & \textbf{All-CNN} & \textbf{Blk-CNN} & \textbf{W-POD} & \textbf{FW-CNN} \\
\midrule
\endhead
\midrule
\multicolumn{7}{r}{{Continued on next page}} \\
\bottomrule
\endfoot
\bottomrule
\endlastfoot
\texttt{v25\_alpha2\_norm} & 0.9970 & 0.9970 & 0.9965 & 0.9959 & 0.9973 & 0.9956 \\
\texttt{v25\_alpha\_norm} & 0.9967 & 0.9967 & 0.9963 & 0.9960 & 0.9971 & 0.9949 \\
\texttt{v25\_mean\_norm} & 0.9840 & 0.9840 & 0.9881 & 0.9877 & 0.9839 & 0.9870 \\
\texttt{v25\_sin2\_norm} & 0.9971 & 0.9971 & 0.9965 & 0.9959 & 0.9974 & 0.9956 \\
\texttt{v25\_sin3\_norm} & 0.9966 & 0.9966 & 0.9958 & 0.9948 & 0.9970 & 0.9948 \\
\texttt{v25\_sin\_norm} & 0.9968 & 0.9968 & 0.9963 & 0.9960 & 0.9971 & 0.9950 \\
\texttt{v25\_sincos\_norm} & 0.9966 & 0.9966 & 0.9960 & 0.9957 & 0.9970 & 0.9946 \\
\texttt{v2\_alpha2\_norm} & 0.9972 & 0.9972 & 0.9967 & 0.9962 & 0.9975 & 0.9959 \\
\texttt{v2\_alpha\_norm} & 0.9970 & 0.9970 & 0.9965 & 0.9962 & 0.9973 & 0.9952 \\
\texttt{v2\_cos2\_norm} & 0.9856 & 0.9856 & 0.9888 & 0.9880 & 0.9856 & 0.9877 \\
\texttt{v2\_mean\_norm} & 0.7208 & 0.7208 & 0.6976 & 0.7766 & 0.7289 & 0.7937 \\
\texttt{v2\_sin2\_norm} & 0.9972 & 0.9972 & 0.9967 & 0.9960 & 0.9975 & 0.9957 \\
\texttt{v2\_sin3\_norm} & 0.9967 & 0.9967 & 0.9959 & 0.9948 & 0.9971 & 0.9949 \\
\texttt{v2\_sin\_norm} & 0.9970 & 0.9970 & 0.9965 & 0.9962 & 0.9973 & 0.9953 \\
\texttt{v2\_sincos\_norm} & 0.9969 & 0.9969 & 0.9962 & 0.9960 & 0.9973 & 0.9949 \\
\texttt{v3\_alpha2\_norm} & 0.9968 & 0.9968 & 0.9963 & 0.9956 & 0.9972 & 0.9954 \\
\texttt{v3\_alpha\_norm} & 0.9964 & 0.9964 & 0.9960 & 0.9955 & 0.9968 & 0.9944 \\
\texttt{v3\_mean\_norm} & 0.9789 & 0.9789 & 0.9792 & 0.9632 & 0.9788 & 0.9577 \\
\texttt{v3\_sin2\_norm} & 0.9969 & 0.9969 & 0.9963 & 0.9956 & 0.9973 & 0.9953 \\
\texttt{v3\_sin3\_norm} & 0.9964 & 0.9964 & 0.9956 & 0.9947 & 0.9969 & 0.9946 \\
\texttt{v3\_sin\_norm} & 0.9965 & 0.9965 & 0.9960 & 0.9955 & 0.9969 & 0.9944 \\
\texttt{v3\_sincos\_norm} & 0.9963 & 0.9963 & 0.9958 & 0.9952 & 0.9968 & 0.9941 \\
\texttt{v\_mean\_norm} & 0.9806 & 0.9806 & 0.9912 & 0.9910 & 0.9802 & 0.9913 \\
\end{longtable}

\end{document}